\documentclass{aastex631}

\usepackage{booktabs}
\begin{document}

\title{\texttt{AXP4}, a Numerical Simulator of Pulse Profiles for Binary Accreting X-ray Pulsars -- I: \\ Beam Geometries}

\correspondingauthor{Parisee S. Shirke}
\email{parisee@iucaa.in}

\author[0000-0003-2977-3042]{Parisee S. Shirke}
\affiliation{Inter-University Centre for Astronomy and Astrophysics,\\
Post Bag 4, Ganeshkhind, \\
Pune 411 007, India}

\author[0000-0001-8604-5362]{Swarnim S. Shirke}
\affiliation{Inter-University Centre for Astronomy and Astrophysics,\\
Post Bag 4, Ganeshkhind, \\
Pune 411 007, India}

\begin{abstract}
The paper presents model pulse profiles from binary accreting X-ray pulsars using high-resolution numerical simulations for pencil and fan emission beam geometries, each with two different optical depths obtained using a new numerical code \lq \texttt{AXP4}' developed in this work. Sharp, pointed-peak improvements are obtained in the known flat-top pulse profiles by extrapolating the model emission functions available in the literature over the full range of emission angles for a realistic description. The effect of general relativistic bending of light due to the strong gravity near the neutron star is taken into account, updating previous attempts with new analytic approximations for photon trajectories in curved space-time reported in the literature (including additional flux contribution from the dark \lq unseen' face of the pulsar). Incorporating the other general relativistic effect of gravitational red-shift is seen to redden the pulsar beam to increasingly lower energies along the X-ray observing band. Integrating this emission over a circular slab geometry for two antipodal hotspots lying flush at the neutron star polar caps presents smoothened, composite gravitationally bent pulse profiles to the observer. A comparison between the corresponding non-normalized flux profiles is included. For theoretical completeness and reference base, this treatment is applied to isotropic emitters with uniform injection and limb-darkened beams, as well. The possibility of the production of Einstein rings around a pulsar for a limiting value of the compactness parameter is a distinctive feature. 
\end{abstract}


\section{Introduction} \label{sec:Intro}

Due to a combination of a variety of unique characteristics, namely a strong magnetic field strength $\sim$10$^{12}$ G, high temperatures $\sim$10$^7$ K and a high matter density, X-ray pulsars are rich physical and astrophysical laboratories and promising targets for X-ray space missions. When a strongly magnetized neutron star in a binary system accretes matter from its companion star, the in-falling matter gets threaded onto the magnetic field of the neutron star and is funneled onto its magnetic poles creating hotspots. The gravitational potential energy is dissipated away as hot thermal Brem\ss trahlung X-ray radiation. With a typical luminosity of 10$^{37}$ ergs/s, such binary accreting X-ray pulsars (XRPs) form the category of some of the brightest galactic X-ray sources in the high-energy sky. The persistent bright emission from high mass binary accreting X-ray pulsars can exploited for examining their pulse phase-resolved characteristics.

Since the magnetic pressure is far stronger than the radiation and gas pressure, it is the dominant factor in governing the direction of X-ray propagation. In the presence of a strong magnetic field around the neutron star, the outgoing X-radiation gets collimated along a preferential direction of propagation, due to magneto-scattering with the accreted plasma matter. The result is an emission \lq beam' emanating from the polar caps. As a pulsar rotates about an axis misaligned with the magnetic dipole axis, its emission beams sweep across the observer's line-of-sight, producing pulsating measurements of X-ray flux for an observer on Earth, as though a galactic X-ray \lq light-house'. When the anisotropic beams sweep across the observer's line-of-sight, the emission is detected as periodic, coherent \lq pulses' from the target source in the X-ray band. The observed X-ray pulse profiles are investigated in the paper improving previous attempts. Several of these are simulated for various source configurations by the custom, generalized \lq \texttt{Accreting X-ray Pulsar Pulse Profile Producer' (AXP4)} code developed in this work. 

The extent of beam collimation is characterized by the \lq pulsed fraction' of the observed pulse profile which is a standard physical quantity measuring the difference between the maximum and the minimum value of X-ray flux escaping along the most and the least preferred emission angles, respectively. For low mass accretion, the outgoing emission from circular polar \lq hotspot' slab regions is collimated into a conical \lq pencil' beam along the direction of the magnetic dipole axis. High rates of accretion can be sustained by binary accreting X-ray pulsars in high mass X-ray binaries, which result in the formation of a column of accreted plasma \citep{1975A&A....39..185I, 1976MNRAS.175..395B, 1981A&A....93..255W, 2015MNRAS.447.1847M, 2021MNRAS.501..564G}, that can support super-Eddington luminosity through outward radiation pressure. Matter free-falls beneath a collisionless radiation-dominated shock front formed above the neutron star surface after being shocked to lower velocities. Radiation-matter interactions in the magnetically-confined column \citep{Becker&Wolff2005, Becker&Wolff2007} cause the X-ray photons to escape from the sides of the accretion column perpendicular to the direction of the magnetic dipole axis producing a \lq fan' beam. In the former case, the observer detects maximum X-ray flux when the hotspot points toward the observer and in the latter case, the flux peaks when the emission region points perpendicular to the line-of-sight as radiation escapes perpendicular to the magnetic axis. A combination of the two is also quite possible e.g. the high mass X-ray binary, Centaurus X-3 observed in \cite{2021JApA...42...58S} is considered a prime candidate for a mixed pencil (slab) and fan (column) emission geometry \citep{1997ApJ...477..897B, 2024A&A...687A.210L} (In particular, a switch between the two mechanisms can be observed as a phase shift in the pulse profile). The pencil and fan beams remain observationally degenerate. Resolving the observational degeneracy between the two models is an open problem, motivating the need for geometry-based simulations and the ongoing inclusion of X-ray polarimetric instruments in global space missions like \textit{XPoSat} \citep{Rishin2010}, \textit{IXPE} \citep{IXPE}, \textit{XIPE} \citep{XIPE}, etc. Numerical simulation efforts like the ones presented in this paper, although polarisation-summed, are expected to take a theoretical step toward resolving this degeneracy. 

The distribution of X-ray flux with the emission angle measured from the direction of the magnetic axis of the pulsar is a quantitative description of the emission beam. Such a beaming function characterizes the shape of the outgoing X-ray beam whether pencil for slabs or fan for columns. Different beam patterns give rise to different pulse profiles and can arise from different emission geometries and thus, point to different accretion scenarios \citep{GnedinSunyaev1973}. Different accretion processes give rise to different accretion geometries (See \cite{Becker2012} and the works of \citeauthor{2022arXiv220414185M}) namely filled column, hollow rings \citep{2013MNRAS.433.3048K}, accretion mounds \citep{Dipanjan2012, 2020MNRAS.497.1029B}. Crescent-shaped structures \citep{miller2019psr} and multiple oval spots \citep{2019ApJ...887L..21R, miller2019psr} have been reported recently. Combinations of these geometries are also a likely possibility \citep{2019ApJ...887L..21R}, thus yielding various hotspot configurations. Instabilities can give rise to stochastic fluctuations in the shape of the accretion region \citep{1980A&A....90..359H}, even producing fractional rings. In numerical modeling studies, the beaming pattern directly determines the pulse shape dependence on the spin phase of the rotating pulsar and vice versa for data-driven observational studies. Using the radiative transfer computations by \citet{MeszarosNagel1985a} for pencil and fan emission, the expected emergent characteristics were computed by \citet{MeszarosNagel1985b} for a set of geometrical configurations that was restricted in range.  

While direct extended, high-resolution imaging studies are not possible for \lq point-source' compact stars, the rotation of the neutron star which produces a periodic variation in the observed flux can be probed with timing studies. The resulting observations are in the form of a periodic pulse profile. The high energy emission produces pulses that seldom fall to zero intensity and exhibit a large duty cycle $(\geq 50\%)$ distinguishing the pulses of X-ray pulsars from radio pulsars. The pulse shape typically appears with a primary peak corresponding to the more visible emission region and a secondary inter-pulse from the second emission region (See e.g. \cite{2021JApA...42...58S}). X-ray pulsar pulse profiles are seen to vary with the energy\footnote{This can result from the higher radiation fanning out more or from absorption due to non-spherical symmtery} and luminosity, the latter being correlated with the geometry of emission region \citep{1973A&A....25..233G, 10.1111/j.1365-2966.2008.13251.x, 2015MNRAS.448.2175L, 2020MNRAS.491.1857D} for different accretion rates. Studies over a wide range of luminosities can provide insight into the range of corresponding mass accretion rates that govern resultant emission geometries. This has a corresponding effect on spectral and beam formation as observed in the resultant pulse profile. 

While the majority of studies of X-ray pulsars have been observational and phenomenological so far (See e.g. studies with the \textit{UHURU, Tenna, Ginga, OSO-8, Ariel, Einstein, CHANDRA, XMM-Newton, NICER,
NuSTAR, AstroSat, eROSITA}, etc. missions), any attempts towards ground-up numerical modeling to predict such observations by employing a sound theoretical understanding of the underlying physical mechanisms and the interplay of various contributing factors are few and far between and can be updated with better computing facilities available today (See \cite{Meszaros1988, 2021MNRAS.501..109C} and references therein). To add to the difficulty, the theory of pulse profile formation is far more intricate than the simplified models typically used (See \cite{mushtukov2022accreting} for a review). A realistic prediction for individual pulsars on a case-to-case basis would require the full knowledge of the isolated and combined effects of several inter-dependent factors \citep{GnedinSunyaev1973} like the accretion rate, magnetic field strength and configuration, emission geometry and beam, X-ray luminosity and energy, pulsar geometry, observer's line-of-sight and particularly, the effect of gravitational light bending \citep{RiffertMeszaros1988, 2001ApJ...563..289K, PoutanenBeloborodov2006, Mushtukov2015} with the added possibility of multiple and additional sources of emission \citep{2019ApJ...887L..21R, 2015MNRAS.448.2175L, 2018MNRAS.474.5425M}, which can only be best attempted with a piece-wise bootstrapping approach (See the works of \citeauthor{mushtukov2022accreting}). Furthermore, there are more possible advanced physical effects enlisted in \cite{2021MNRAS.501..109C, 2021MNRAS.501..129C} that may also come into play. 

The inclusion of light bending can significantly increase the pulsed fraction in a fan beam and affect the correlation between actual and apparent luminosity \citep{2020PASJ...72...34I, 2018MNRAS.474.5425M}.
While \cite{MeszarosNagel1985b}'s profiles have been studied with light bending in \cite{RiffertMeszaros1988} for isotropic emission functions, \cite{Meszaros1988} and \citep{mushtukov2022accreting} identify improved simulations including gravitational light bending effects as a focus area for better modeling X-ray pulsar characteristics. The bending of X-ray photon trajectories emanating from the pulsar as they traverse through the strong gravitational field around the neutron star before reaching the observer is the primary effect to be taken into consideration for neutron stars and black holes (See e.g. \cite{Beloborodov2002}, \cite{PoutanenBeloborodov2006}, \cite{Falco2016} and \cite{Dipanjan2012}). Failure to do so can result in a gross mismatch between the observed flux of X-rays and that predicted by radiative transfer theory (See \cite{kraus1998light} for a review). \cite{2020A&A...640A..24P} have summarized analytic approximations for photon trajectories in a Schwarzschild space-time metric. While all formulae offer marginal improvements for finer modeling, they all provide a sufficiently robust description of curved photon trajectories, for practical purposes. This paper puts together the independent, dedicated advancements in the theory of gravitational light bending with high energy emission from strongly magnetized neutron stars and thereby, presents model gravitationally bent \citeauthor{MeszarosNagel1985b} profiles using \citeauthor{Beloborodov2002}'s approximation for the full emission angular range at high phase resolution. The \lq \texttt{GR Light Bender}' module so developed can handle all possible source inclinations. In keeping with \citeauthor{2019ApJ...887L..21R} and \citeauthor{miller2021radius}'s reports of slab hotspots, composite hotspot-integrated profiles are also included by developing a \lq \texttt{Composite Slab Integrator}' module. Isotropic and limb-darkened injections encountered in the literature \citep{RiffertMeszaros1988, MeszarosNagel1985b} are further included for comparison and theoretical completeness.

Recently, \citeauthor{mushtukov2022accreting} have systematically compiled and outlined various aspects of X-ray pulsars in a review article. Having explored some of the topics of coherent pulsations, mass transfer in binary systems, accretion flows and spin-up/down mentioned therein, in an earlier paper \citep{2021JApA...42...58S}, we investigate the pulse profiles and the geometry and physics of emitting regions in this paper (See \cite{tauris2023physics} for a subject review). Additionally, noted here are, firstly, the insignificance of any non-accretion component of X-ray luminosity due to off-axis thermal contribution from nuclear burning and secondly, the effect of the geometry of emitting region on beam pattern formation and its relation with the mass accretion rate which thereby, along with luminosity, becomes an implicit parameter in the numerical simulations presented in this paper. 

Work in this paper is proposed to be devoted to a final model and detailed investigation of beamed emission in accreting binary X-ray pulsars and building a custom tool to predict these observables for a wide range of physical conditions at high resolution to obtain updated, refined estimates of the initial numerical models for high mass X-ray binaries. Observables for different geometries, defined by the relative angles between the spin axis, the magnetic axis and the view axis can be computed. The resulting simulated light curves can be directly compared with observations from the existing and upcoming missions (see accompanying paper, \cite{2024arXiv240807504S}). 

\newpage
\subsection{Assumptions}

\subsubsection{Model dipolar magnetic field}
To proceed with the numerical implementation, a standard dipolar pulsar magnetic field of $\sim$$10^{12}$ G (higher quadrupoles of the magnetic moment are insignificant at larger scales \citep{long2007accretion}) is assumed. This naturally leads to the next assumption of model antipodal emission regions, thus paving the way for a textbook cornerstone exercise, which can be later scaled up in the future to accommodate advanced effects required to model more realistic peculiarities which can deviate from model behaviour. 

While the assumption might provide generalized robustness, the occurrences of non-dipolar and distorted magnetic field structures have been widely reported in the literature \citep{1996ApJ...457L..85K, 2017Sci...355..817I, 2017A&A...605A..39T, 2022MNRAS.515..571M} and also confirmed in \cite{2021JApA...42...58S}, which can be accompanied by several different hotspot configurations as well (See e.g. \cite{2019ApJ...887L..21R}). \cite{Dipanjan2012} have discussed how the local magnetic field lines can get distorted due to accumulation at the edges of an accretion mound on the polar caps of the neutron star. The magnetic field strength can also vary over time through different physical mechanisms like Ohmic decay, Hall effect \citep{2021Univ....7..351I} and accretion processes \citep{Mukherjee2017}). (See \cite{konar1997magnetic} for a review).

\subsubsection{Circular slab emission geometry}
The work primarily uses a circular slab as in \cite{MeszarosNagel1985a} as a cornerstone exercise that can be scaled to model other, realistic, more intricate geometries. These correspond to other different possible beaming configurations, namely, (i) isotropic emitter, (ii) isotropic emitting limb-darkened slab/column, (iii) cylindrical column \citep{MeszarosNagel1985a}, (iv) hollow column \citep{2001ApJ...563..289K}, (v) conical column and (vi) mound \citep{Dipanjan2012, 2020MNRAS.497.1029B}, (vii) multiple oval spots \citep{2019ApJ...887L..21R, miller2019psr} and (viii) filled crescents \citep{miller2019psr}. Discussions on point isotropic emitters, isotropic slabs, isotropic fans and point fans emanating from hypothetical columns are included to a limited extent, for the sake of comparison and completeness. In the case of a fan, for optically thick emission, the radiation emanates from the walls of the accretion column and for optically thin emission, the radiation can stream freely from the polar slab itself.

\subsubsection{Homogeneous atmosphere}
This is the scattering environment produced by the accreted hot, tenuous plasma -- assumed to be homogeneous, mostly ionized hydrogen with trace helium -- falling onto the polar caps which Comptonizes outgoing thermal X-rays \citep{1979A&A....78...53B}.
The plasma is more transparent parallel to the direction of the magnetic field than perpendicular to it for a pencil beam emanating from a slab geometry and vice versa for a fan beam from column geometry. Three different values of temperatures are co-existent in the plasma, namely (i) the ion temperature, (ii) the electron temperature parallel to the magnetic field, $\vec{B}$ and (iii) the temperature representing the electron distribution over the Landau levels, thus resulting in non-thermal local equilibrium (non-LTE) conditions.

The following sections in the paper describe the numerical simulations of model pulse profiles for binary accreting X-ray pulsars by building on previous attempts in the literature (See e.g. \cite{1981A&A...102...97W, 1991MNRAS.251..203L} and similar works based on \cite{MeszarosNagel1985a}, e.g. \cite{RiffertMeszaros1988, 1993ApJ...406..185R, 1994A&A...286..497P, 1995MNRAS.277.1177L, 2001ApJ...563..289K, 2004ApJ...613..517L, 2007ESASP.622..403T}). Specifically experimenting with the prescription by \cite{MeszarosNagel1985a}, a robust simulator code is developed which can account for two geometries of the emission region namely, slab and column, corresponding to both beaming geometries namely, pencil beam and fan beam, respectively, each for two different optical depths. This provides four different geometric configurations namely, shallow slab, deep slab, shallow column, and deep column kept consistent with \cite{MeszarosNagel1985a}. Since the emission geometry is governed by the accretion rate and luminosity, the two naturally get included as implicit parameters in the code. For each case, 8 discrete values of X-ray photon energies are used. These are suitably chosen in keeping with \citeauthor{MeszarosNagel1985b} such that they span the full $1-100$ keV range. Unlike \citeauthor{MeszarosNagel1985a}'s limited consideration of specific combinations of inclinations of the rotation axis and magnetic axis of the pulsar with the observer's line of sight, the code has been generalized to handle any arbitrary angular inclination. 

The section-wise layout of the paper provides the detailed module-wise numerical scheme implemented in developing the \lq \texttt{Accreting X-ray Pulsar Pulse Profile Producer' (AXP4)} code. After this Introduction in Section \ref{sec:Intro}, details of the physical set-up for the undertaken numerical implementation and simulator code operation are provided in Section \ref{sec:methods}.

In Section \ref{subsec:beamed}, the model emission functions provided in Figs. 7 and 8 of \cite{MeszarosNagel1985b} (by performing radiative transfer of X-ray photons in model pulsar atmospheres with Feautrier discrete ordinate method) for two antipodal emission points are used. These model emission beams of binary accreting X-ray pulsars serve as the input to the code for reproducing the numerical simulations of flat-top pulse profiles to verify the correctness of code operation. For a more realistic consideration, the angular range of \citeauthor{MeszarosNagel1985a}'s emission beams (chosen using a Gaussian quadrature scheme suitable for their radiative transfer) is completed over the full possible range of emission directions by extrapolating these curves and using geometric reasoning, to finally obtain model pointed-peak profiles for binary accreting X-ray pulsars in Figs. \ref{fig:pulses_extrapolated_1} and \ref{fig:pulses_extrapolated_2} at a high spin phase resolution of the pulsar. 

These model pulse profiles traverse in bent trajectories through the strong gravitational field around the neutron star before reaching the observer. Sub-section \ref{gbl} describes the \lq \texttt{GR Light Bender}' numerical module developed in the now \texttt{AXP4} code. Based on recent advances in the theory of curved photon trajectories in a Schwarzschild metric around a neutron star, updated, corrected profiles at high resolution are provided in Figs. \ref{fig:pulses_gbl_1} and \ref{fig:pulses_gbl_2}, which are in good agreement with earlier attempts in this direction. Additional flux received from the unseen face of the neutron star is also modeled. The other relevant general relativistic effect of gravitational red-shift is also taken into account. 

In Section \ref{hotspot_int}, the simulations are further improved by developing another new module, \lq \texttt{Composite Slab Integrator}' which breaks \citeauthor{MeszarosNagel1985a}'s assumption of point emission regions and accounts for a more realistic geometry of hotspots by integrating the emission over a circular slab at the polar caps. Such comprehensive composite gravitationally bent and hotspot-integrated pulses are also simulated at high resolution. For the sake of theoretical completeness, this study is further extended in the accompanying paper, \cite{2024arXiv240807504S} to examine the effect of an increasing slab area (to geometrically spherical caps on the neutron star) on the pulsed fraction, for the full span of surface coverage (0\%--100\%) at high resolution with a new, custom \lq \texttt{Surface Coverage}' module (cylindrical surfaces can be constructed for accretion columns). Modified pulse profiles for circular slabs are thus obtained by integrating the emission over two antipodal circular slabs lying flush on the neutron star surface at the antipodal polar caps, with \citeauthor{MeszarosNagel1985a}'s point emission functions used for modeling the emission from each infinitesimal surface area element. The two modules so developed can be run individually or in conjunction with each other. The resultant model composite surface-integrated and relativistically modified profile simulations are reported in Figs. \ref{fig:pulses_surfint}. For comparison, we supplement these profiles generated using model emission beams with alternative pulse profiles generated using hypothetical, fiducial input beams from uniform, isotropic test emitters in Section \ref{app:isotropic}. 

 New results obtained with this code and subsequent discussions are presented in Section \ref{sec:rnd}, followed by the key conclusions summarized in Section \ref{sec:conclusions}. 

\section{Numerical Simulations}\label{sec:methods}

A numerical simulator is developed to grow realistic pulse profiles for XRP astronomical sources. The tool can handle various input parameters corresponding to model physical conditions for binary accreting X-ray pulsars which are prime targets for X-ray observations. This is achieved using direct and exact modeling of pulsar rotation for a dipolar model magnetic field of $\sim$$10^{12}$ Gauss and 8 discrete values of X-ray frequencies spanning the 1 -- 100 keV range. Any arbitrary combination of values of source and observer inclinations can be handled by the code.

\cite{MeszarosNagel1985a, MeszarosNagel1985b} have previously performed radiative transfer in pulsar atmospheres for X-ray emission spanning $1-100$ keV.
Their calculations assume a static, homogeneous neutron star atmosphere with purely thermal sources and no external illumination. The magnetic field lines are perpendicular to the emission surface in the slab and parallel to the emitting sides i.e. along the axis, in the column. 
Using $kT$ = 8 keV, $\hbar \omega_B$ = 38 keV, and $\rho$ = 0.5 g/cm$^3$ as model plasma parameters, a standard emission area of about a few times $10^{10}$ cm$^2$ provides a typical X-ray luminosity of $10^{37}$ ergs/s. Their simulations deal with both shallow and deep geometries corresponding to optical depths with $y=5 \times 10^1$ g/cm$^3$ and 
$y=5 \times 10^4$ g/cm$^3$, respectively for a linear depth of $R=10^2$ cm (Thomson optical depth $\tau_T= 2 \times 10^1$) and $R=10^5$ cm (Thomson optical depth $\tau_T = 2 \times 10^4$), respectively. 

The result of their radiative transfer is the polarisation-summed total differential flux $F_{\omega}(\theta) = I_{\omega}(\theta) \cos \theta$ for a circular slab and $F_{\omega}(\theta) = I_{\omega}(\theta) \sin \theta$ for a cylindrical column, with the trigonometric factors chosen to produce an appropriate limb darkening effect. These are seen in Figs. 7 and 8 of \cite{MeszarosNagel1985b}. The flux is a function of the emission angle $\theta$ made by the photon direction of propagation $\vec{k}$ with the magnetic axis $\vec{B}$. A restricted range of eight different emission angles $\theta$ = 88.8$^{\circ}$, 84.2$^{\circ}$, 76.3$^{\circ}$, 65.9$^{\circ}$, 53.7$^{\circ}$, 40.3$^{\circ}$, 26.1$^{\circ}$, 11.4$^{\circ}$ are sampled (using a Gaussian quadrature scheme for determining the discrete values of the angular grid, with $\mu_i = \cos \theta_i$ as the independent variable distributed between 0 and 1), to minimize numerical error. The radiative transfer is performed for a set of eight discrete values of photon energies $\omega = 1.6, 3.8, 9.0, 18.4, 29.1, 38.4, 51.7, 84.7$ keV spanning the 1 -- 100 keV energy range around the medium X-ray band. Since $\tau_T >>1$ in both cases, the curves for the shallow and deep geometries differ only slightly in detail and not drastically in their overall qualitative nature. The curve for 38.4 keV is closest to the cyclotron energy ($\hbar \omega_B$ = 38 keV) and exhibits a distinct nature (See \cite{MeszarosNagel1985b} for a detailed discussion of the underlying physical interpretation).

\section{Beamed Pulse Profiles} \label{subsec:beamed}

In this sub-section, \citeauthor{MeszarosNagel1985b}'s model emission beams for binary accreting X-ray pulsars (Figs. 7 and 8 of \cite{MeszarosNagel1985b}) are first used to reproduce \citeauthor{MeszarosNagel1985b}'s model pulse profiles (Figs. 9 and 10 of \cite{MeszarosNagel1985b} and its Sec. IV for a detailed discussion of the pulse phase-dependent features), by numerically implementing their direct calculation of the pulsar rotation through geometric considerations of spherical trigonometry. The emission beam shapes from Figs. 7 and 8 of \cite{MeszarosNagel1985b} are re-used\footnote{\url{https://plotdigitizer.com/}} as they provide the emitted intensity as a function of emission angles, $\theta$ within the X-ray beam (pencil/fan) of the pulsar. As described in Section \ref{sec:Intro} a high spin phase-resolution \texttt{AXP4} code is developed for this purpose. The numerical tool can handle any arbitrary set of inclination angles $(i_1, i_2)$ for 8 discrete X-ray photon energies $\times$ 2 hotspot geometries $\times$ 2 optical depths. Building on these, full pointed-peak simulations obtained by extrapolating \citeauthor{MeszarosNagel1985b}'s limited angular range to the full angular range of emission beams of $[0\textsuperscript{o}, 90\textsuperscript{o}]$ are presented in Figs. \ref{fig:pulses_extrapolated_1} and \ref{fig:pulses_extrapolated_2}.

\begin{figure}
    \centering
    \includegraphics[clip, trim = 240 25 180 35, width=.8\textwidth]{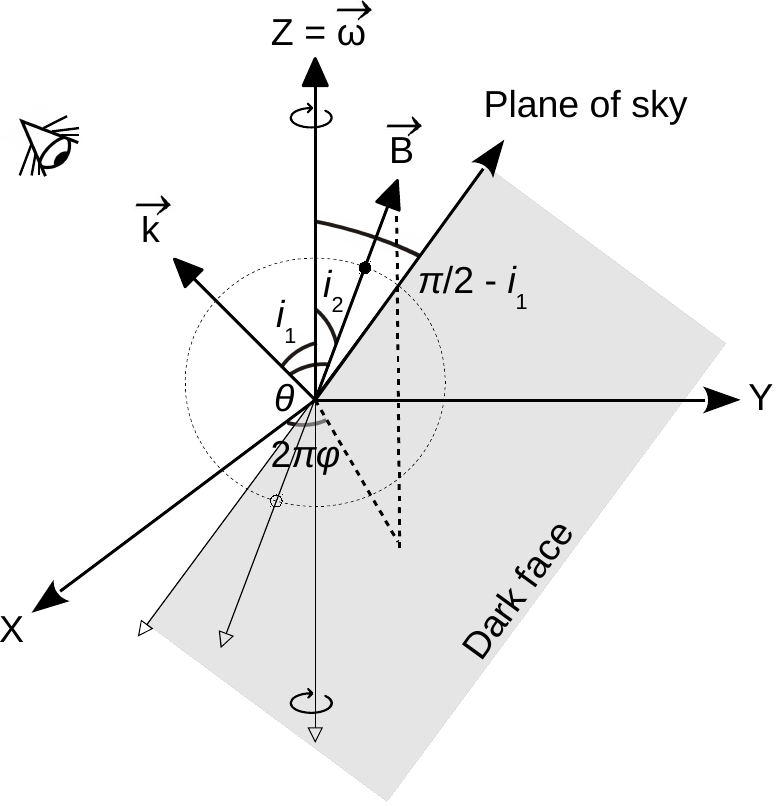}
    \caption{Sketch of the geometrical configuration of a model pulsar. The co-ordinate system is set up with its origin at the center of the neutron star and the $Z$-axis along the pulsar rotation axis, $\vec{\omega}$. The direction of the observer's line of sight is the same as the photon propagation direction, $\vec{k}$. $i_1$ and $i_2$ are the inclination angles of the observer's line of sight and the magnetic dipole axis, $\vec{B}$ with respect to the rotation axis of the pulsar. The angle $\theta$ made by the magnetic axis with the observer's line of sight is spin phase-dependent and can be obtained using the trigonometric relation between these angles using Eq. (\ref{rotation}). The spin phase is measured by $2\pi\phi$ which is the angle made by the projection of the magnetic axis on the $XY$-plane. The $X$-axis is chosen to lie in the plane defined by $\vec{\omega}$ and $\vec{k}$ and the direction of the $Y$-axis is determined by its orthogonality with the $X$- and $Z$-axes. The sky plane is perpendicular to the observer's line of sight and separates the visible front face of the neutron star from its invisible back \lq dark\rq~face. The dots represent the location of the emission regions coincident with the base of the magnetic dipole axis on the neutron star surface, the solid one being the visible emission area and the hollow one being the antipodal emission area located on the unseen face. (Refer also, Fig. 1(a) from \cite{Meszaros1988}).}
    \label{fig:pulsar_geometry}
\end{figure}

\citeauthor{MeszarosNagel1985b}'s limited choice of inclination combinations is maintained in this paper, namely, (i) (50$^{\circ}$, 20$^{\circ}$) where $(i_1+i_2) < 90^{\circ}$, (ii) (45$^{\circ}$, 45$^{\circ}$) where $(i_1+i_2) = 90^{\circ}$ and $|i_1-i_2|$ = 0$^{\circ}$ and (iii) (75$^{\circ}$, 45$^{\circ}$), (60$^{\circ}$, 45$^{\circ}$) and (80$^{\circ}$, 60$^{\circ}$) were ($i_1+i_2$) $> 90^{\circ}$ to facilitate a direct comparison. As the reproduced profiles seem to be in exact agreement, the code is further generalized to handle all possible values of angles of inclination possible for binary accreting X-ray pulsars, i.e., $i_1$, $i_2 \in [0\textsuperscript{o},~180\textsuperscript{o}]$.

The beamed nature of the radiation emitted by a rotating pulsar along with a misalignment between the rotation and magnetic axes, gives rise to pulses as seen by the observer as given by the exact formula through spherical trigonometric considerations,
\begin{equation} \label{rotation}
    \cos{\theta}=\cos{i_1}\cos{i_2}+\sin{i_1}\sin{i_2}\cos{2\pi\phi},
\end{equation}
where $\theta$ is the angle between the magnetic dipole axis and the observer's line of sight, $i_1$ is the inclination of the observer's line of sight (also the direction of the wave vector, $\vec{k}$) from the rotation axis, ($\vec{\omega}$), $i_2$ is the inclination of the magnetic dipole axis ($\vec{B}$) with the rotation axis and $\phi$ is the spin phase as shown in Fig. \ref{fig:pulsar_geometry}.

$\phi$ parametrizes the spin phase of pulsar rotation with \texttt{0.0} corresponding to a face-on view, \texttt{0.5} for a face-off view and \texttt{0.25} and \texttt{0.75} for side views. $\theta$ provides the angle of the polar cap (coincident with the root of the magnetic axis on the surface of the neutron star) with the observer's line of sight as a function of the pulsar spin phase. The angles of inclination of the observer's line of sight ($i_1$) and that of the magnetic axis ($i_2$) with respect to the rotation axis of the pulsar should lie within the range [0\textsuperscript{o}, 180\textsuperscript{o}]. As $\phi$ assumes values in the range $[0, 0.5]$, $\theta$ varies in the range $[|i_1-i_2|, i_1+i_2]$ and when $(i_1+i_2)$ exceeds $90\textsuperscript{o}$, $\theta'=180\textsuperscript{o}-\theta$ for the second hotspot which comes into view. This expression is symmetric in $i_1$ and $i_2$. For slabs, the pulse peaks at $\phi=\texttt{0.0}$ and for cylinders, at $\phi=\texttt{0.25}$\footnote{Breaking the observational degeneracy between these models is an open problem. Resolution of this degeneracy would be readily possible with the dawning advent of X-ray polarimetric studies. The phase-resolved sweep of the polarisation vector is expected to display very distinct behaviour for the two geometries.}. When $(i_1+i_2) > 90$\textsuperscript{o}, an inter-pulse is expected from the second pole coming into view. 

\subsection{Extrapolation of beams over the full emission angular range} \label{extrap}
The pulsar beams provided by \cite{MeszarosNagel1985b} are further standardized by extending the available range of emission angles -- limited due to the use of a Gaussian quadrature scheme required in \citeauthor{MeszarosNagel1985b}'s radiative transfer computations -- to span the full possible range of [0\textsuperscript{o}, 90\textsuperscript{o}] for model and improved reference. The \texttt{numpy.interp()} function from the NumPy library\footnote{\url{https://numpy.org}} \citep{harris} used to interpolate the beaming function for an arbitrary emission angle, provides flat values outside the upper and lower limits of a data range which gives \lq flat-top' pulse profiles when extreme values are sampled (See e.g. (45\textsuperscript{o}, 45\textsuperscript{o}) case for shallow slab). Replacing this with \texttt{scipy.interpolate.interp1d()} from the SciPy library\footnote{\url{https://scipy.org}} \citep{virtanen} provides extrapolation outside the available data range sharpening the \lq flat-tops' to \lq pointed-peaks' as seen in Figs. \ref{fig:pulses_extrapolated_1} and \ref{fig:pulses_extrapolated_2}.

Figs. \ref{fig:pulses_extrapolated_1} and \ref{fig:pulses_extrapolated_2} show the normalized pulse profiles with extrapolation to cover the full range of beam emission directions from 0$^{\circ}-90^{\circ}$~for shallow and deep slabs and columns for five orientations of viewing angles ($i_1$, $i_2$) from left to right, (50$^{\circ}$, 20$^{\circ}$), (45$^{\circ}$, 45$^{\circ}$), (75$^{\circ}$, 45$^{\circ}$), (60$^{\circ}$, 45$^{\circ}$) and (80$^{\circ}$, 60$^{\circ}$), with $i_1$ and $i_2$ being inter-changeable. These are shown for the X-ray photon energies 1.6, 3.8, 9.0, 18.4, 29.1, 38.6, 51.7 and 84.7 keV in a staggered manner by adding unity in each case. The profiles have been reproduced as per the prescription given by \cite{MeszarosNagel1985b} at a finer phase resolution of $0.001$. Phase \texttt{0.0} corresponds to a face-on view of the emission region and phase \texttt{1.0} corresponds to the same after the completion of one rotation spin of the pulsar. The profiles are in good agreement with Figs. 9 and 10 of \cite{MeszarosNagel1985b}, with the extrapolation providing sharpened pointed-peaks for the (45$^{\circ}$, 45$^{\circ}$) orientation.

\subsection{Trigonometric boundary conditions} \label{sec:pulse_pro_trig}
As the differential flux $F_{\omega}(\theta) = I_{\omega}(\theta) \cos \theta$ for a circular slab and $F_{\omega}(\theta) = I_{\omega}(\theta) \sin \theta$ for a cylindrical column is presented with trigonometric factors chosen to produce an appropriate limb darkening effect, trigonometric boundary conditions of $F_{\omega}(\theta=90\textsuperscript{o}) = I_{\omega}(\theta) \cos(\theta=90\textsuperscript{o}) = 0$ and $F_{\omega}(\theta=0\textsuperscript{o}) = I_{\omega}(\theta) \sin( \theta=0\textsuperscript{o})=0$ can be imposed in theory. A distinct notch-dip is seen for columns for the (45\textsuperscript{o}, 45\textsuperscript{o}) orientation at spin phase \texttt{0.0} that samples close to the boundary and similarly, a sharpening of the pulse peak is seen for the same geometry for slabs.

\section{Gravitationally modified pulse profiles} \label{gbl}

The X-ray photons traversing through the strong gravitational field around a neutron star do not travel toward the observer in a straight Cartesian line. The observed energy of the photons is also slightly different from their as-emitted energy. This section discusses the primary general relativistic modifications 
to the intrinsic beamed simulated pulse profiles for better modeling of observed profiles. Model gravitationally bent pulse profile simulations, (i) using \citeauthor{Beloborodov2002}'s analytic approximation for photon trajectories in strong gravity, (ii) including additional gravitationally bent contribution from the dark face of the neutron star and (iii) using the \texttt{GR Light Bender} module developed in this work are presented in Figs. \ref{fig:pulses_gbl_1} and \ref{fig:pulses_gbl_2}. The panels from Figs. \ref{fig:pulses_extrapolated_1} and \ref{fig:pulses_extrapolated_2} now correspond to new values of gravitationally red-shifted X-ray photon energies as shown in Table \ref{tab:z}. 

\begin{figure}
    \centering
    \includegraphics[clip, trim = 140 30 140 30, width=\textwidth]{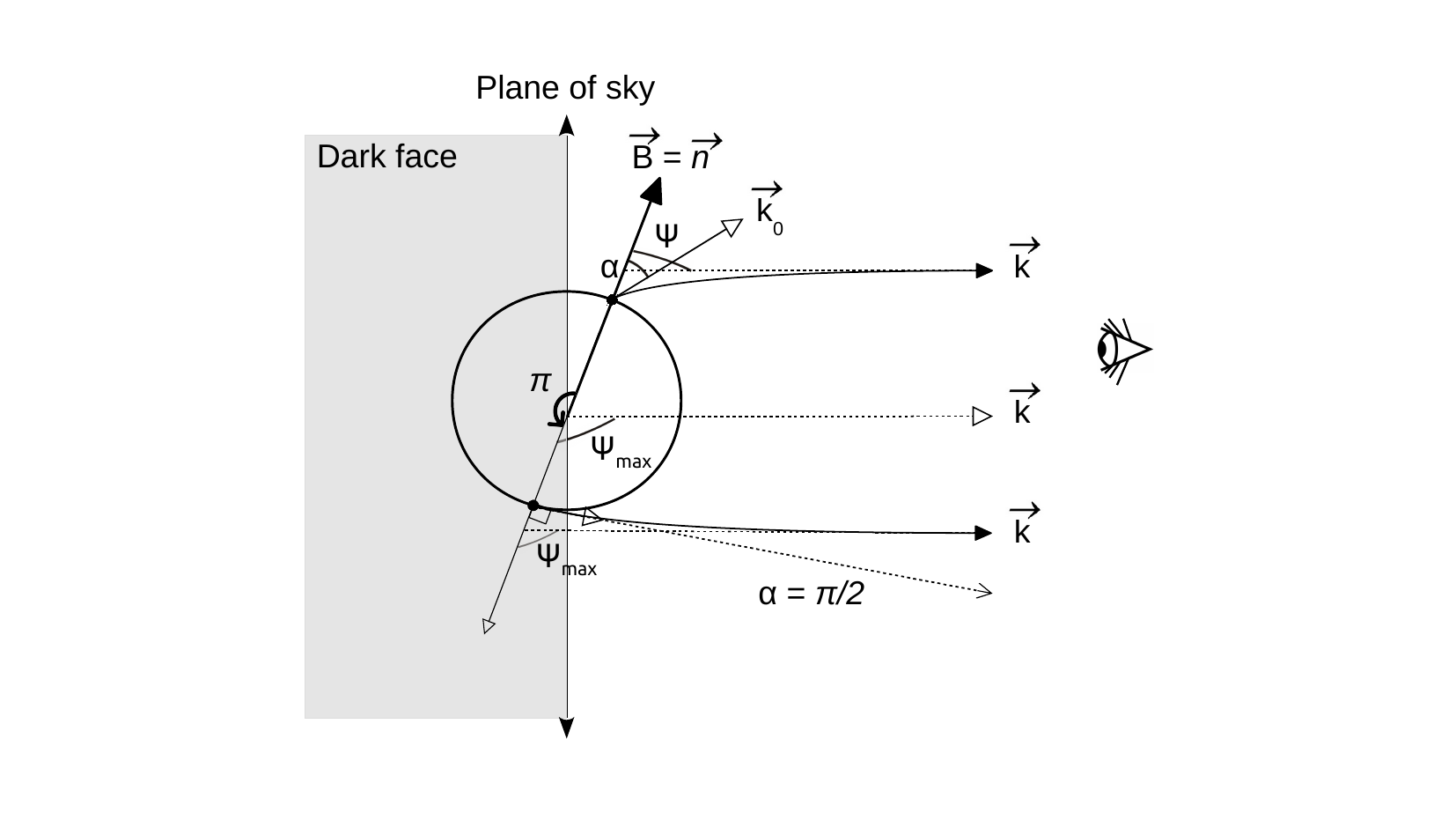}
    \caption{Sketch of the geometric configuration of gravitationally bent X-ray photon trajectories while traversing through the strong gravitational field of the source pulsar. The plane of the sky is perpendicular to the observer's line of sight, $\vec{k}$. The magnetic axis, $\vec{B}$ is along the normal to the surface, $\vec{n}$ at the emission location. X-rays emitted at an initial direction $\vec{k_0}$ at an initial beaming angle, $\alpha$ get bent to a final direction $\vec{k}$ making a new angle $\psi$ with the magnetic axis for an observer located at infinity. The total viewing range now gets extended from being limited to the emission range of $[0\textsuperscript{o}, 90\textsuperscript{o}]$ to a gravitationally bent range of $[0,\psi_{\text{max}}]$. Accordingly, an antipodal emission region located on the invisible back \lq dark\rq~face of the neutron star can contribute additional, bent X-ray flux, especially prominent for a fan beam. (See also, Fig. B2 from \cite{Dipanjan2012} and Fig. 1 from \cite{PoutanenBeloborodov2006}.)}
    \label{fig:diagram_gbl}
\end{figure}

\subsection{Photon trajectories in curved space-time}
\label{Metho:2}

The \citeauthor{Beloborodov2002}'s approximation for bending of light due to gravity is given as:
\begin{equation} 
    \cos{\alpha}=u+(1-u)\cos{\psi}
    \label{beloborodov}
\end{equation}
where $\alpha$ is the local emission angle measured with respect to the local normal to the emission surface and $\psi$ is the angle between the radius vector of the position of the emission region and the observer's line of sight as shown in Fig. \ref{fig:diagram_gbl}. The compactness parameter $u=r_{\text{Sch}}/R_{NS}$ is the Schwarzschild radius of a typical neutron star in units of radius  $\sim$10 km \citep{Beloborodov2002, PoutanenBeloborodov2006}. For a model neutron star with a mass of 1.4 M$_{\odot}$ and a radius of 10 km \citep{bhattacharya1991formation}, the compactness parameter $u=0.413$.

This is a straight-forward but approximate equation used to relate the angles of initial launching and the final ray asymptote towards the observer through the strength of the gravitational field i.e. the curvature of traversed space-time. It is valid for trajectories beyond two Schwarzschild radii ($r>2r_{\text{Sch}}$) from the center of the gravitational potential well. This holds well for binary accreting X-ray pulsars, where the emission occurs from the polar caps $r>3 r_{\text{Sch}}$ (See Appendix \ref{exact}). If the error is defined as $e=\delta \xi/ \xi$ with $\xi=\psi-\alpha$, then \citeauthor{Beloborodov2002}'s approximation is accurate to an error of $1\%$ within $\alpha<75\textsuperscript{o}$ which goes to up to $3\%$ for the maximum value of  $\alpha=90\textsuperscript{o}$ \citep{phdthesisSandeep}.

\subsection{Scheme}

The pulse profiles (without accounting for gravitational light bending, i.e. $\alpha=\psi=\theta$) are seen in Figs. \ref{fig:pulses_extrapolated_1} \& \ref{fig:pulses_extrapolated_2} from the beaming functions in Figs. 7 and 8 of \cite{MeszarosNagel1985b}
The change while accounting for light bending is that the direction ($\theta$) corresponding to the line-of-sight of an observer at infinity ($\psi$) and the photon emission angle ($\alpha$) are now two different directions related by an analytic formula (Eq. (\ref{beloborodov})) measuring the relation between the two through general relativistic considerations. The additional contribution from the dark face of the neutron star must also be taken into account. This brings extra radiation -- which would otherwise not be received -- into the observer's field of view due to strong gravity. In mathematical terms:\\
Putting $\theta = \psi$ in Eq. (\ref{rotation}),
\begin{equation}
    \cos{(\theta = \psi)}=f(i_1, i_2, \phi).
\end{equation}
We can calculate corresponding $\alpha$ using Eq. (\ref{beloborodov}),
\begin{equation}
    \cos{\alpha}=g(\psi),
\end{equation}
from which the corresponding emission intensity can be read off using $\theta=\alpha$ in the beaming function in Figs. 7 and 8 of \citep{MeszarosNagel1985b} where,
\begin{equation}
    I=h(\theta=\alpha).
\end{equation}

\subsubsection{Additional contribution from the dark face} \label{secondhotspot}

A distinct feature of strong gravitational bending around compact stars is the extra flux received from the back \lq dark' face of the star which would otherwise have not asymptoted towards the observer in flat space-time. As shown in Fig. \ref{fig:diagram_gbl}, emission from the second antipodal hotspot gets additionally bent towards the visible half of the neutron star containing the first hotspot. In mathematical terms, $\alpha \in [0\textsuperscript{o},~90\textsuperscript{o}]$ constraints $\psi$ to vary in the range $[0\textsuperscript{o}, \psi_{\text{max}}]$, where $\psi_{\text{max}}=\cos^{-1}\Big({\frac{-u}{1-u}}\Big)$ (from Eq. (\ref{beloborodov})). Using the value of $u$ estimated above, this yields $\psi \in [0\textsuperscript{o}, 134.817\textsuperscript{o}]$. Thus, we sub-divide the range for $\theta$ obtained from Eq. (\ref{rotation}) into three cases namely, (i) Single hotspot only: $\theta \in [0\textsuperscript{o}, 180\textsuperscript{o}-\psi_{\text{max}}] \sim [0\textsuperscript{o}, 45.183\textsuperscript{o}]$. Only one hotspot is visible and the prescription in Sec. \ref{gbl} is executed as is. (ii) Both hotspots: $\theta \in [180\textsuperscript{o}-\psi_{\text{max}}, \psi_{\text{max}}] \sim [45.183\textsuperscript{o}, 134.817\textsuperscript{o}]$. Emission from both the hotspots is received by the observer as a result of gravitational bending. Thus, Sec. \ref{gbl} must be executed for both the hotspots with $\theta'=180\textsuperscript{o}-\theta$ for the second hotspot and the contributions thus computed individually and then, added. (iii) Second hotspot only: $\theta \in [\psi_{\text{max}}, 180\textsuperscript{o}] \sim [134.817\textsuperscript{o}, 180\textsuperscript{o}]$. Sec. \ref{gbl} must be executed with $\theta'=180\textsuperscript{o}-\theta$ for the second hotspot alone. Results are shown in Figs. \ref{fig:pulses_gbl_1} and \ref{fig:pulses_gbl_2}.

\subsection{Trigonometric boundary conditions}

Similar to the previous Sec. \ref{sec:pulse_pro_trig}, a distinct notch-dip is seen for columns for the (45\textsuperscript{o}, 45\textsuperscript{o}) orientation at spin phase \texttt{0.0} that samples close to the boundary and similarly, a sharpening of the pulse peak is seen for the same geometry for slabs. Since such sharp discontinuities may not be representative of a realistic scenario, these profiles are not presented in this paper.

\subsection{Gravitational red-shift}

Radiation emanating from regions of strong gravity loses energy in overcoming a strong gravitational pull. The resulting general relativistic stretching in wavelength is called gravitational red-shift. The extent of energy loss depends on (i) the strength of the gravitational field characterized by the Schwarzschild radius, $r_s = 2GM/c^2$ and (ii) the depth of the region where the radiation originated within the gravitational potential well, $R_e$\footnote{When the photon does not originate at the gravitationally \lq lensing' compact star, the concerned physical quantity would be the \lq impact parameter' which is the closest distance of photon approach measured from the center of the neutron star.} (equal for all photons in a pencil beam generated on a slab). The overall red-shift effect is characterized by the $z$ parameter given by the standard formula,
\begin{equation}
    z  = \frac{\Delta \lambda}{\lambda_e} = \frac{GM}{c^2 R_e} = \frac{r_s}{2 R_e}~,
\end{equation}
where, $\lambda_e$ denotes the emitted wavelength and $\Delta \lambda$ measures the difference between the observed wavelength ($\lambda_o$) and $\lambda_e$. Suitably re-arranging the terms in the equation above gives,
\begin{equation}
    \lambda_o = \lambda_e \Bigg[ \Bigg( \frac{r_s}{2R_e}  \Bigg) + 1 \Bigg] = \lambda_e \Bigg[ \Bigg( \frac{u}{2}  \Bigg) + 1 \Bigg].
\end{equation}
where, $u = r_s/R_e$ is the compactness parameter.

Using the reciprocal nature of photon energy ($E$) and corresponding radiation wavelength ($\lambda$), we arrive at another standard relation,

\begin{equation} \label{redshift}
    E_o = \frac{E_e}{(1+z)}
\end{equation}
where, $z = r_s/2R_e$ with $r_s$ being the Schwarzschild radius for a model binary accreting X-ray pulsar with mass $M_{\text{NS}}$ = 1.4 M$_{\odot}$, measured from the center of the neutron star and $R_e$ being the distance from the center at which the X-rays originate. Since the emission comes from the hotspot located at the polar caps, $R_e = R_{\text{NS}}$ = 10 km. In terms of the compactness parameter, $u=r_s/R_e$, $z=u/2$.

A {third GR effect is that of a Shapiro time delay between the pulse arrival time from the two antipodal hotspots due to the different photon trajectories traversed by each in a strong gravitational potential well. This effect, however, is found to be minor in practice and is expected to produce a small phase delay of a few percent in fast-rotating millisecond pulsars, at best. (See Eq. (19) and the discussion following Eq. (10) in \cite{2020A&A...640A..24P}). A fourth GR effect is that of frame dragging which applies to spinning black holes and not the slow-rotating neutron stars here. The special relativistic modifications of Doppler boosting/beaming and relativistic aberration apply largely to fast-rotating millisecond pulsars. Additional relativistic effects arise due to caustics and the presence of an accretion disk (higher order images, self-irradiation) \citep{dovciak2010probing} but remain irrelevant for the scope of this work.

Figs. \ref{fig:pulses_gbl_1} and \ref{fig:pulses_gbl_2} show the normalized gravitationally bent pulse profiles including contribution from the dark face for a model neutron star of 1.4 M$_{\odot}$ mass and 10 km radius covering the full extrapolated range of beam emission directions from 0$^{\circ}$ -- 90$^{\circ}$~for shallow and deep, slab and column for the same five orientations of viewing angles (50$^{\circ}$, 20$^{\circ}$), (45$^{\circ}$, 45$^{\circ}$), (75$^{\circ}$, 45$^{\circ}$), (60$^{\circ}$, 45$^{\circ}$), (80$^{\circ}$, 60$^{\circ}$), with $i_1$ and $i_2$ being inter-changeable. These are shown for the corresponding red-shifted X-ray photon energies 1.3, 3.1, 7.5, 15.2, 24.1, 32.0, 42.8 and 70.2 keV as per Table \ref{tab:z} in a staggered manner by adding unity in each case. The profiles have been obtained by subjecting \citeauthor{MeszarosNagel1985b}'s procedure to \citeauthor{Beloborodov2002}'s analytic light bending approximation, at a finer phase resolution of \texttt{0.001}. The profiles exhibit general relativistic modifications while remaining consistent with \citeauthor{MeszarosNagel1985b}'s results. By visual inspection, the generated pulse profiles seem to be broadened as the observer now sees closer to the magnetic axis due to bent photon trajectories. 

\section{Composite Slab Integration}\label{hotspot_int}

While \cite{RiffertMeszaros1988} have reported column-integrated bolometric pulse profiles including isotropic injection and there have been other reports discussing polar cap coverage by circular slabs, this section presents model slab-integrated bolometric pulse profiles for model \citeauthor{MeszarosNagel1985b} beaming functions for both shallow and deep geometries. Composite model slab integrated and relativistically modified pulse profile simulations -- derived by a double summation over a discretized grid representing a circular slab at the polar cap and by performing a suitable rotation of the co-ordinate to add a \texttt{Composite Slab Integrator} module in the \texttt{AXP4} code -- are presented in Fig. \ref{fig:pulses_surfint}. 

\begin{figure}
    \centering 
    \includegraphics[clip, trim = 0 10 0 10, width=\textwidth]{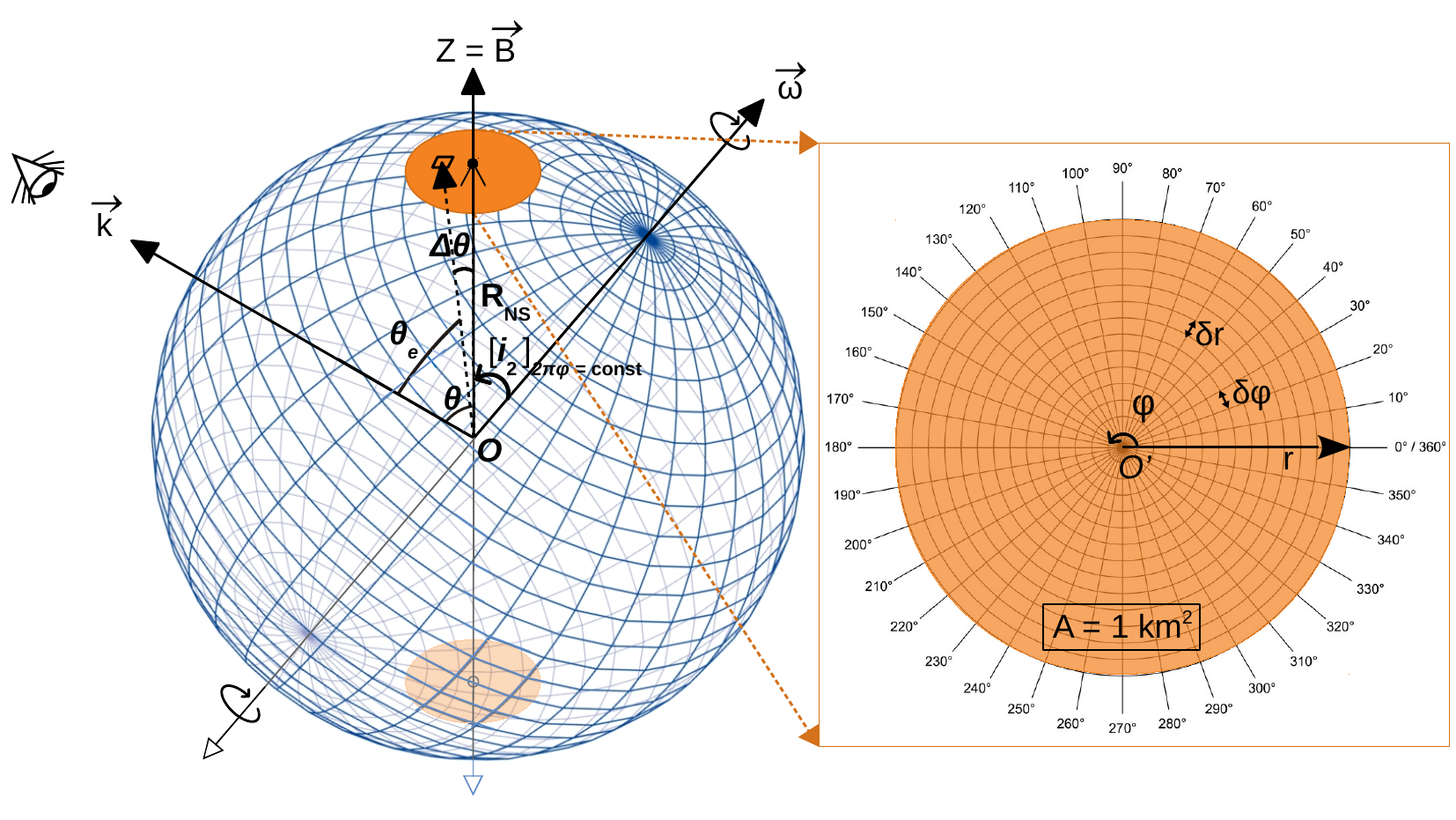}
    \caption{Sketch of the geometric construction of a finite polar grid $(r, \phi)$ on model antipodal circular slabs lying at the polar caps of a pulsar with a standard emission area of 1 km$^2$ and shown in orange colour. The overall pulsar geometry characterized by $\vec{k}, \vec{B}$ and $\vec{\omega}$ and the corresponding angles are the same as shown in Fig. \ref{fig:pulsar_geometry}. The $Z$-axis is rotated by an angle $i_2$ keeping the value of the spin phase constant. $\Delta \theta$ is the slab-resolved polar angle of an arbitrary infinitesimally small area element and $R_{\text{NS}}$ is the radius of the neutron star (= 10 km, for a model pulsar). In this transformed frame of reference, $\theta$ is geometrically similar to $i_1$ in Fig. \ref{fig:pulsar_geometry}, $\Delta\theta = r/R_{NS}$ is similar to $i_2$, $\theta_{e}$ corresponds to $\theta$ and $\phi$ corresponds to $2\pi\phi$ and are substituted in Eq. (\ref{rotation}). \textit{(inset)} $(\delta r, \delta \phi)$ represent the polar grid cell resolution.}
     \label{fig:diagram_surf_int}
\end{figure}

\subsection{Double integral over a closed surface}

The projected intensity of emission from the beaming function is given as $I\cos{\theta}$ where the trigonometric cosine accounts for the angle between the emission direction and the normal to the emitting surface, $S$. Then, the flux $F$ received from an anisotropic radiator is given as a double integral over two closed circular surfaces as slabs.
The discretized summation would run over all area elements that contribute X-radiation towards the observer at infinity. This would also include those area elements from the dark face of the neutron star that contribute extra flux at the detector through the gravitational bending of X-rays.

\newpage
\subsection{Polar co-ordinates for the construction of a flat discretized grid}

To numerically implement this surface integration, the polar caps are represented by two model circles divided into a discrete polar grid of an arbitrary resolution. For two model circular hotspots lying flush on the neutron surface, the surface can be suitably represented by a polar $(r,\phi)$ Boyer-Lindquist co-ordinate system \citep{dovciak2010probing}, where $r$ represents the radial distance from the center of the pole and $\phi$ represents the azimuthal angle measured from an arbitrary direction as shown in Fig. \ref{fig:diagram_surf_int}. Surface integration would require a summation of the flux from all infinitesimal area elements $\Delta A_{r_i, \phi_j}, \forall~i, j$ in a polar co-ordinate system centered at the magnetic pole \citep{Dipanjan2012} (See also, \citep{PoutanenGierlinski2003}).

When $\Delta A$ is resolved into two-dimensional co-ordinates, Eq. (\ref{flux}) contains $r \cos{\alpha}$ as a suitable Jacobian factor, $\mathcal{J}$ based on the emission geometry (circular slab/cylindrical column) and thereby, the co-ordinate system of choice, \citep{Dipanjan2012} (polar/cylindrical) as the local surface area element $dS (r,\phi)$ gets distorted as it gets transformed from the Boyer-Lindquist source plane to the detector plane.

Assuming a standard emission area of a single hotspot to be $\sim$1~km$^2$, the sum can be computed iteratively by suitably varying the polar and azimuthal co-ordinates as, 

\begin{equation} \label{flux}
     F= \sum\limits_{r=0}^{0.564 \text{ km}} \sum\limits_{\phi=0\textsuperscript{o}}^{360\textsuperscript{o}} I(\alpha) \cos{\alpha}~r \Delta r \Delta \phi. 
\end{equation}

where $I(\alpha)\cos{\alpha}$ is used from the beaming function in Figs. 7 and 8 from \cite{MeszarosNagel1985b}, the infinitesimal area element $dS = r dr d\phi$ for a visible emission element and the polar Jacobian, $\mathcal{J} = r$.

\subsubsection{Grid Resolution}
The resolution chosen for the grid is $\Delta r=0.01$ km in keeping with the resolution of NICER results \citep{miller2021radius}. The azimuthal resolution is taken as $\Delta \phi = 10\textsuperscript{o}$, dividing the grid into 36 angular divisions about the center of the polar cap. This provides a hotspot resolution of $\Delta A_{r,\phi}|_{r=0.56 \text{ km}}$$\approx$$0.001$ km$^2$ near the outer circumference of the hotspot with a finer resolution towards the inner regions.

\subsection{Rotation transformation of co-ordinate axes}
In Eq. (\ref{rotation}), the angles of the hotspot $\theta$ with respect to the observer are described completely, using angles of inclination $i_1$ and $i_2$ that measure the angle of the location of the center of the hotspot and observer's line of sight with respect to the rotation axis of the pulsar. $\phi$ is allowed to vary fully from $[0,1]$ with $2\pi\phi$ measuring the phase within such a full rotation spin. However, with a circular slab, there is an extended area of emission around the center of the hotspot. Since we consider model circular hotspots, these are modeled by two perfect circular patches around the central location described by Eq. (\ref{rotation}) with their normals $\hat{n}$ coinciding with the direction of the magnetic field.

\subsubsection{Differential emission angle}

Within the double summation, it is the angle of the infinitesimal area element $\theta_{\text{element}}$ which is required instead of $\theta_{\text{center}}$. This sum must be carried out over all radial and azimuthal co-ordinates within the closed hotspot surface. The X-rays emanating from each infinitesimally small area element on the polar grid will get bent slightly differently depending on the local launching angle $\alpha_{\text{element}}$ which asymptotes towards the observer for any given sight-line for a fixed set of inclinations $(i_1, i_2)$.

A clever trick in transforming the co-ordinate axes and bringing a corresponding transformation in the co-ordinates under consideration can make such a calculation possible using the same analytic relation between similar quantities as those in Eq. (\ref{rotation}). For this, the co-ordinates axes are rotated keeping the center of the system fixed in its place, i.e. the new and old origins remain the same. The $Z$-axis is rotated such it changes from being coincident with the rotation axis, $\vec{\omega}$ to being coincident with the magnetic axis, $\vec{B}$. Since the location of the hotspot is coincident with the base of the magnetic field on the neutron star surface, it lies in the same plane which contains the curve describing the rotation of the $Z$-direction on the neutron star. The location of the center of the polar cap is described by its distance from the center of the new co-ordinate system of magnitude $R_{\text{NS}}$ with vanishing angular co-ordinates.

Now an area element on the polar grid in the vicinity of such a point can be treated as an \lq old'~point hotspot offset from the center of the $Z$-axis by an angle $\Delta\theta = r/R_{NS}$. The observer's line of sight would need to be defined with respect to the new axis - this is $\theta_{\text{element}}(i_1, i_2)$ calculated for an arbitrary pulsar inclination using Eq. (\ref{rotation}). As the summation runs over both the polar co-ordinates, $\Delta\theta$ varies in the range $[0, r_{\text{hotspot}}/R_{\text{NS}}]$ and the azimuthal co-ordinate varies within $[0,2\pi]$ thus, corresponding to the full rotation spin of the old point hotspot. Substituting these variables in Eq. (\ref{rotation}), yields the angle made by the observer's line of sight with the location of the infinitesimal area element measured from the center of the neutron star. Thus, using this resemblance with the previous prescription, Eq. (\ref{rotation}) can be re-written to compute $\theta_{\text{element}}$ for each infinitesimal area element as, 

\begin{equation} \label{area_rotation}
    \cos{\theta_{\text{element}}}=\cos{\theta}\cos{\Delta\theta}+\sin{\theta}\sin{\Delta\theta}\cos{\phi}~.
\end{equation}

For a given $\theta$, as $r$ varies, we calculate $\Delta \theta$ as $\Delta \theta  \in \Big[ 0, r/R_{NS}\Big] \text{i.e.} \Big[ 0, \frac{0.5 \text{km}}{10 \text{km}}\Big]$ so that $\theta$ varies in the range $[|\theta- \Delta \theta|, \theta + \Delta \theta]$. Using the following resemblance with the previous prescription, we can use Eq. (\ref{rotation}) again, to compute $\theta_{\text{element}}$ for each infinitesimal area element. Treating $\theta_{\text{element}}$ as $\psi_{\text{element}}$, $\alpha_{\text{element}}$ can be derived from Eq. (\ref{beloborodov}) and $I(\alpha)\cos{\alpha}$ can be obtained from the beam function for each specific area element.
Contribution from the second hotspot can be added as outlined in Sec. \ref{secondhotspot}. Final simulations are shown in Fig. \ref{fig:pulses_surfint}.

\section{Results and Discussions} \label{sec:rnd}

\subsection{From flat-top to pointed-peak beams}

The \lq model' pulse profiles for pencil beams presented in \cite{MeszarosNagel1985b} appear with flat tops near spin phase $\phi$ = \texttt{0.0} when the pulsar points straight toward the observer. However, this was found to be a direct consequence of the limited range of emission angles available with the Gaussian quadrature scheme used for minimizing the numerical error while performing radiative transfer to arrive at the source photon densities. In a realistic scenario, all angles between $\alpha$ = 0\textsuperscript{o} to 90\textsuperscript{o} are available for emission, with the effective beaming direction being governed by interactions with the strong pulsar magnetic field. Using linear extrapolation, the beam functions are extended to cover this full range of available angles. Trigonometric considerations also present usable boundary condition alternatives. Introducing these replaces the apparent flat-tops (formed by directly joining emission from the lowest available angles by a flat line) with sharp, pointed peaks by allowing the profiles to sample emission from directions close to spin phase $\phi$ = \texttt{0.0}, especially for a pencil beam. Even the dips seen in the second panel from the bottom in Fig. \ref{fig:pulses_extrapolated_1} for a $(45\textsuperscript{o}, 45\textsuperscript{o})$ slab, are sharper after allowing the beaming range to be extrapolated. Such sharp profiles were reported by \cite{RiffertMeszaros1988} before although in a slightly different exercise. The profiles need to be obtained at a very fine phase resolution to see sharp features/kinks as shown in Fig. \ref{fig:resolution} -- a lower resolution smoothens over them but is sufficient for capturing all the major features. This is especially important to bear in mind with more effects being added in the subsequent sections that can produce fine kinks mid-profile. 

\begin{figure}
        \centering \hfill
        \includegraphics[trim=5 8 5 5, clip=true, height=0.45\textheight]{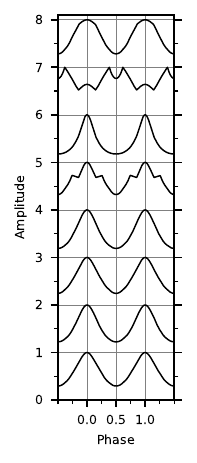}
        \hspace{-.3cm}
        \includegraphics[trim=25 8 5 5, clip=true, height=0.45\textheight]{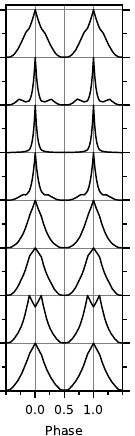}
        \hspace{-.3cm}
        \includegraphics[trim=25 8 5 5, clip=true, height=0.45\textheight]{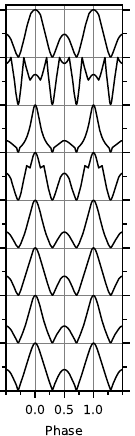}
        \hspace{-.3cm}
        \includegraphics[trim=25 8 5 5, clip=true, height=0.45\textheight]{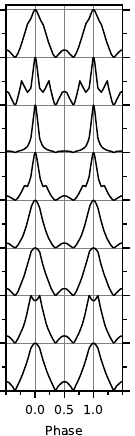}
         \hspace{-.3cm}
        \includegraphics[trim=25 8 5 5, clip=true, height=0.45\textheight]{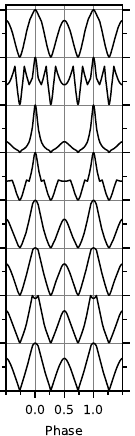} \hfill
         \vspace*{.3cm} \hfill
        \includegraphics[trim=5 8 5 5, clip=true, height=0.45\textheight]{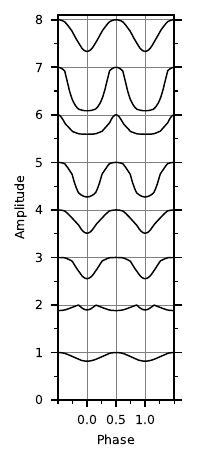}
        \hspace{-.3cm}
        \includegraphics[trim=25 8 5 5, clip=true, height=0.45\textheight]{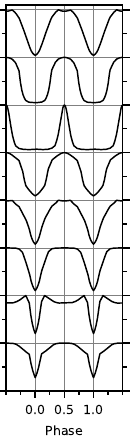}
        \hspace{-.3cm}
        \includegraphics[trim=25 8 5 5, clip=true, height=0.45\textheight]{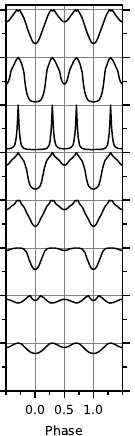}
        \hspace{-.3cm}
        \includegraphics[trim=25 8 5 5, clip=true, height=0.45\textheight]{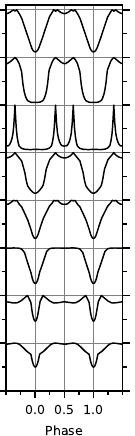}
         \hspace{-.3cm}
        \includegraphics[trim=25 8 5 5, clip=true, height=0.45\textheight]{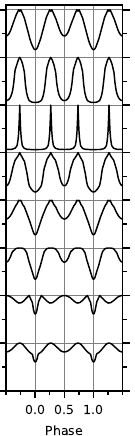} \hfill
        \caption{Normalized pulse profiles with extrapolation to cover the full range of beam emission directions from 0$^{\circ}$ -- 90$^{\circ}$~for shallow (\textit{top}) slab and (\textit{bottom}) column for 5 orientations of viewing angles ($i_1$, $i_2$) from left to right (50$^{\circ}$, 20$^{\circ}$), (45$^{\circ}$, 45$^{\circ}$), (75$^{\circ}$, 45$^{\circ}$), (60$^{\circ}$, 45$^{\circ}$) and (80$^{\circ}$, 60$^{\circ}$), with $i_1$ and $i_2$ being inter-changeable. These are shown for the X-ray photon energies (\textit{bottom to top}) 1.6, 3.8, 9.0, 18.4, 29.1, 38.6, 51.7 and 84.7 keV in a staggered manner by adding unity in each case. The profiles have been reproduced as per the prescription given by \cite{MeszarosNagel1985b} at a finer phase resolution of $0.001$. Phase \texttt{0.0} corresponds to a face-on view of the emission region and phase \texttt{1.0} corresponds to the same after the completion of one rotation spin of the pulsar. The profiles are in good agreement with Fig. 9 of \cite{MeszarosNagel1985b}, with the extrapolation providing sharpened pointed-peaks for the (45$^{\circ}$, 45$^{\circ}$) orientation.}
        \label{fig:pulses_extrapolated_1}
\end{figure}

\begin{figure}
        \centering \hfill
        \includegraphics[trim=5 8 5 5, clip=true, height=0.45\textheight]{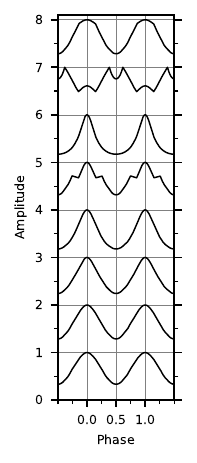}
        \hspace{-.3cm}
        \includegraphics[trim=25 8 5 5, clip=true, height=0.45\textheight]{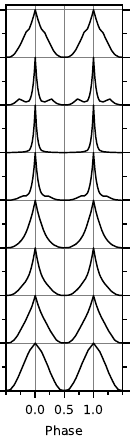}
        \hspace{-.3cm}
        \includegraphics[trim=25 8 5 5, clip=true, height=0.45\textheight]{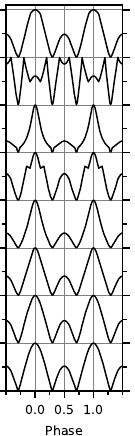}
        \hspace{-.3cm}
        \includegraphics[trim=25 8 5 5, clip=true, height=0.45\textheight]{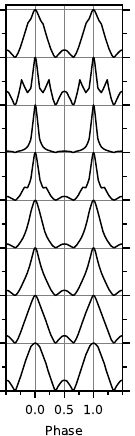}
         \hspace{-.3cm}
        \includegraphics[trim=25 8 5 5, clip=true, height=0.45\textheight]{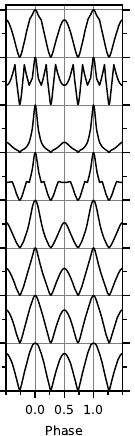} \hfill
         \vspace*{.3cm} \hfill
        \includegraphics[trim=5 8 5 5, clip=true, height=0.45\textheight]{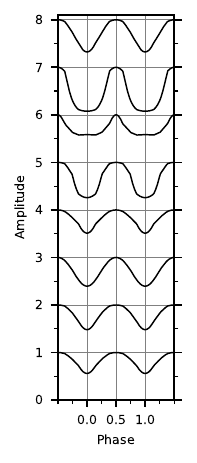}
        \hspace{-.3cm}
        \includegraphics[trim=25 8 5 5, clip=true, height=0.45\textheight]{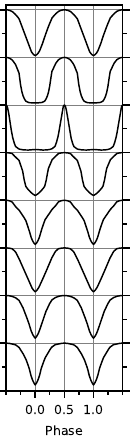}
        \hspace{-.3cm}
        \includegraphics[trim=25 8 5 5, clip=true, height=0.45\textheight]{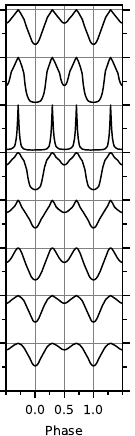}
        \hspace{-.3cm}
        \includegraphics[trim=25 8 5 5, clip=true, height=0.45\textheight]{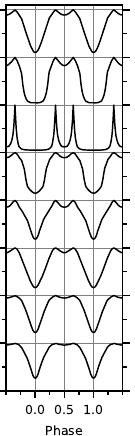}
         \hspace{-.3cm}
        \includegraphics[trim=25 8 5 5, clip=true, height=0.45\textheight]{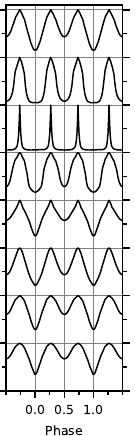} \hfill
        \caption{Normalized pulse profiles with extrapolation to cover the full range of beam emission directions from 0$^{\circ}$ -- 90$^{\circ}$~as in Fig. \ref{fig:pulses_extrapolated_1} for deep (\textit{top}) slab and (\textit{bottom}) column. The profiles remain in good agreement with Fig. 10 of \cite{MeszarosNagel1985b}, with the extrapolation providing sharpened pointed-peaks seen for the (45$^{\circ}$, 45$^{\circ}$) orientation.}
        \label{fig:pulses_extrapolated_2}
\end{figure}

\subsection{General relativistic fan-out of the emission beam including an excess \lq dark' flux}

The primary modification due to the inclusion of general relativistic effects is the stretching out of the pulse profile over the spin phase of the pulsar rotation, as smaller beaming angles get sampled preferentially. For a $(45\textsuperscript{o},45\textsuperscript{o})$ angle of pulsar inclination and shallow geometry shown in Figs. \ref{fig:pulses_gbl_1} and \ref{fig:pulses_gbl_2}, the Compton wings of the main pulse are seen at a phase of \texttt{0.5} (See discussion on \lq peak inversion' in \cite{MeszarosNagel1985b}) giving the impression of an increase in flux far away from the main peak. The overall flux range inc. the non-pulsed component will be higher (or lower) than the gravity-uncorrected counterpart for a pencil beam from a slab (or a fan beam for a column). Radiation directed straight towards the observer will be bent out-of-sight (except $\alpha=\psi=90\textsuperscript{o}$ case). 

The secondary effect is the drastic modification in the simulated pulse profiles before a phase of $\phi$ = \texttt{0.25} (magnetic field $\vec{B}$ lying in the plane of the sky) where extra flux is received within a small phase change due to additional photons bent towards the observer from the unseen face of the neutron star due to photon propagation in a strong gravitational field. This effect is very pronounced for columns, as they have logarithmically high emission near the edges, towards $\alpha$ = 90\textsuperscript{o}; such a strong additional beaming can even overshadow the primary flux seen at phase $\phi$ = \texttt{0.0} as seen in the figure. 

\begin{figure}
        \centering
        \includegraphics[trim=5 8 5 5, clip=true, height=0.45\textheight]{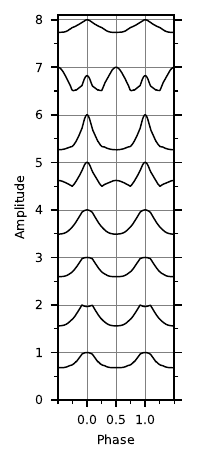}
        \hspace{-.3cm}
        \includegraphics[trim=25 8 5 5, clip=true, height=0.45\textheight]{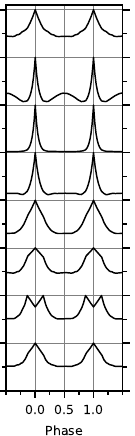}
        \hspace{-.3cm}
        \includegraphics[trim=25 8 5 5, clip=true, height=0.45\textheight]{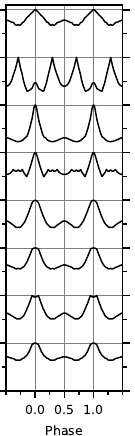}
        \hspace{-.3cm}
        \includegraphics[trim=25 8 5 5, clip=true, height=0.45\textheight]{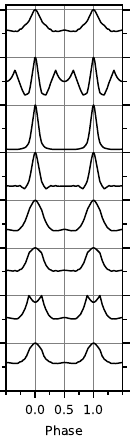}
         \hspace{-.3cm}
        \includegraphics[trim=25 8 5 5, clip=true, height=0.45\textheight]{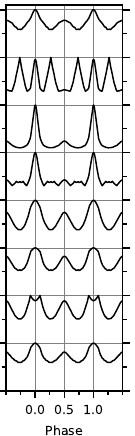}
        \vfill
        \includegraphics[trim=5 8 5 5, clip=true, height=0.45\textheight]{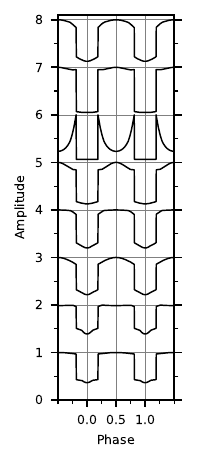}
        \hspace{-.3cm}
        \includegraphics[trim=25 8 5 5, clip=true, height=0.45\textheight]{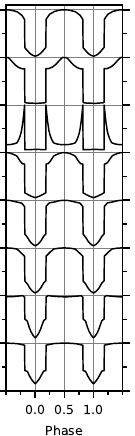}
        \hspace{-.3cm}
        \includegraphics[trim=25 8 5 5, clip=true, height=0.45\textheight]{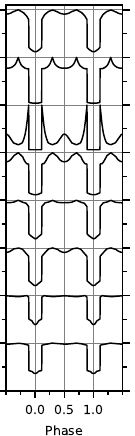}
        \hspace{-.3cm}
        \includegraphics[trim=25 8 5 5, clip=true, height=0.45\textheight]{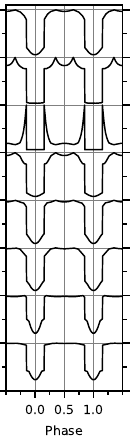}
         \hspace{-.3cm}
        \includegraphics[trim=25 8 5 5, clip=true, height=0.45\textheight]{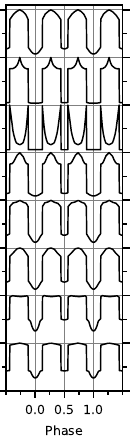}
        \caption{Normalized gravitationally bent pulse profiles including contribution from the dark face for a model neutron star of 1.4 M$_{\odot}$ mass and 10 km radius covering the full range of beam emission directions from 0$^{\circ}$ -- 90$^{\circ}$~for a shallow (\textit{top}) slab and (\textit{bottom}) column as in Figs. \ref{fig:pulses_extrapolated_1} and \ref{fig:pulses_extrapolated_2}. These are shown for the red-shifted X-ray photon energies (\textit{bottom to top}) 1.3, 3.1, 7.5, 15.2, 24.1, 32.0, 42.8 and 70.2 keV in a staggered manner by adding unity in each case. The profiles have been obtained by subjecting \citeauthor{MeszarosNagel1985b}'s procedure to \citeauthor{Beloborodov2002}'s analytic light bending approximation, at a finer phase resolution of $0.001$. The profiles exhibit general relativistic modifications in \citeauthor{MeszarosNagel1985b}'s results.}
        \label{fig:pulses_gbl_1}
    \end{figure}

\begin{figure}
        \centering
        \includegraphics[trim=5 8 5 5, clip=true, height=0.45\textheight]{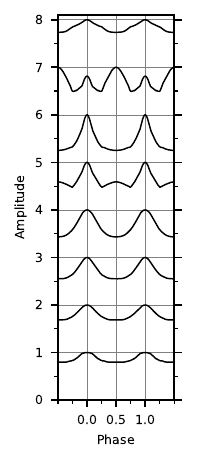}
        \hspace{-.3cm}
        \includegraphics[trim=25 8 5 5, clip=true, height=0.45\textheight]{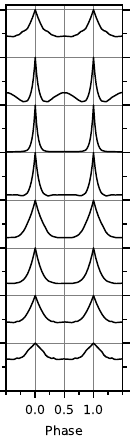}
        \hspace{-.3cm}
        \includegraphics[trim=25 8 5 5, clip=true, height=0.45\textheight]{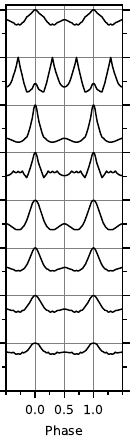}
        \hspace{-.3cm}
        \includegraphics[trim=25 8 5 5, clip=true, height=0.45\textheight]{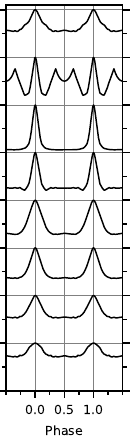}
         \hspace{-.3cm}
        \includegraphics[trim=25 8 5 5, clip=true, height=0.45\textheight]{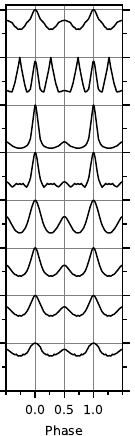}
        \vfill
        \includegraphics[trim=5 8 5 5, clip=true, height=0.45\textheight]{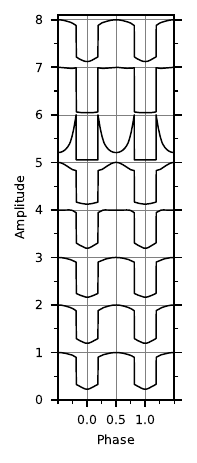}
        \hspace{-.3cm}
        \includegraphics[trim=25 8 5 5, clip=true, height=0.45\textheight]{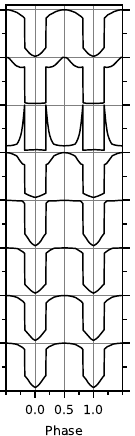}
        \hspace{-.3cm}
        \includegraphics[trim=25 8 5 5, clip=true, height=0.45\textheight]{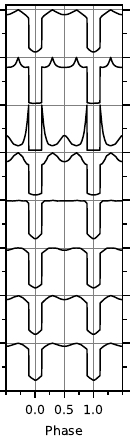}
        \hspace{-.3cm}
        \includegraphics[trim=25 8 5 5, clip=true, height=0.45\textheight]{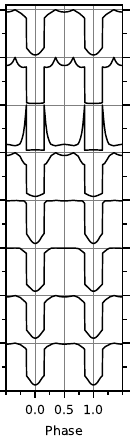}
         \hspace{-.3cm}
        \includegraphics[trim=25 8 5 5, clip=true, height=0.45\textheight]{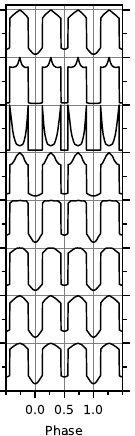}
        \caption{Normalized gravitationally bent pulse profiles including contribution from the dark face for a model neutron star of 1.4 M$_{\odot}$ mass and 10 km radius as in Fig. \ref{fig:pulses_gbl_1}~for a deep (\textit{top}) slab and (\textit{bottom}) column.}
        \label{fig:pulses_gbl_2}
\end{figure}

\subsection{X-ray beam reddening}

The resultant discrete values of X-ray energies no longer match the model beamed \citeauthor{MeszarosNagel1985b} values. The observed values of X-ray photon energies are red-shifted due to the loss of energy in overcoming the strong gravitational pull of the neutron star as per Eq. (\ref{redshift}). The change in X-ray photon energies is summarized in Table \ref{tab:z}. The effect of gravitational red-shift in X-ray photon energy becomes increasingly pronounced for harder X-rays.

\begin{table}
    \centering
    \caption{Discrete values of sampled X-ray photon energies before and after accounting for the effect of general relativistic gravitational red-shift.} 
    \label{tab:z}
    \begin{tabular}{l|cccccccc} \hline\hline
    \multicolumn{9}{c}{X-ray photon energies (keV)} \\ \midrule
        Model \lq$E_e$\rq~$^a$& 1.6 & 3.8 & 9.0 & 18.4 & 29.1 & 38.6 & 51.7 & 84.7 \\
        Red-shifted \lq$E_o$\rq~$^b$& 1.3 & 3.1 & 7.5 & 15.2 & 24.1 & 32.0 & 42.8 & 70.2 \\ \bottomrule \hline 
    \end{tabular}
    \newline 
    \flushleft $^a$ The original values are the same as the model energies from \citeauthor{MeszarosNagel1985b} beaming functions. $^b$ The new values are obtained using Eq. (\ref{redshift}).
\end{table}

\subsection{Formation of an Einstein ring} \label{app:limb}

Since the gravitational bending is a function of the compactness, $u \propto M/R$ the variation in profiles with a varying mass of the neutron star was also studied in Fig. \ref{fig:vary_M}. The radius is computed in each case at a constant density $\rho_{\text{10}} \propto M (=1.4$ M$_{\odot}) / R^3 (=10 \text{ km})$ obtained for a model neutron star. Eliminating the mass from the above two relations gives $ u \propto R^2$ and eliminating the radius gives $u \propto M^{2/3}$. Either way, positive indices indicate an increase in the compactness $u$ with an increase in the mass $M$ and radius $R$ of the neutron star. Depending on the value of mass (and radius), each curve in Fig. \ref{fig:vary_M} would correspond to a different value of gravitational red-shifted X-ray photon energy. Although a constant model neutron star density as calculated for the $M=1.4$~M$_{\odot}, R = 10~\text{km}$ case\footnote{See \cite{1989PASJ...41....1N, _zel_2016} for actual mass and radii distribution of neutron stars.} was assumed as this was sufficient for our exercise in this thesis, do note that this remains a gross approximation, in light of the various neutron star equations of state routinely reported in the literature, arising from several possible distinct neutron star compositions.

The extent of gravitational light bending depends on the compactness $u \propto M/R$ of a compact star. As $u$ increases, an increasing amount of the background hotspot is brought into view by acute bending of photon trajectories. In Fig. \ref{fig:vary_M}, with an increase in the mass of the neutron star, the pulse profiles start flattening out with an increasing extent of gravitational light bending for a limb-darkened beam. The injected beam multiplies the uniform injection by an additional cosine factor appropriate for limb darkening at high angles for the projected slab emission area\footnote{A sine factor would be used for columns.} (See $M=1.4$~M$_{\odot}$ case in Fig. \ref{fig:vary_M}) which further smoothens the input beam. The gravitational bending of emission from the second hotspot pointing away from the observer can produce a full non-pulsed Einstein ring when the bent paths remain very close to the neutron star surface (See e.g. \cite{2021MNRAS.501..109C}). Note that the possibility of the less compact cases representing \textit{pulsed} Einstein rings remains open.

\begin{figure}
    \centering 
    \includegraphics[width=.7\textwidth]{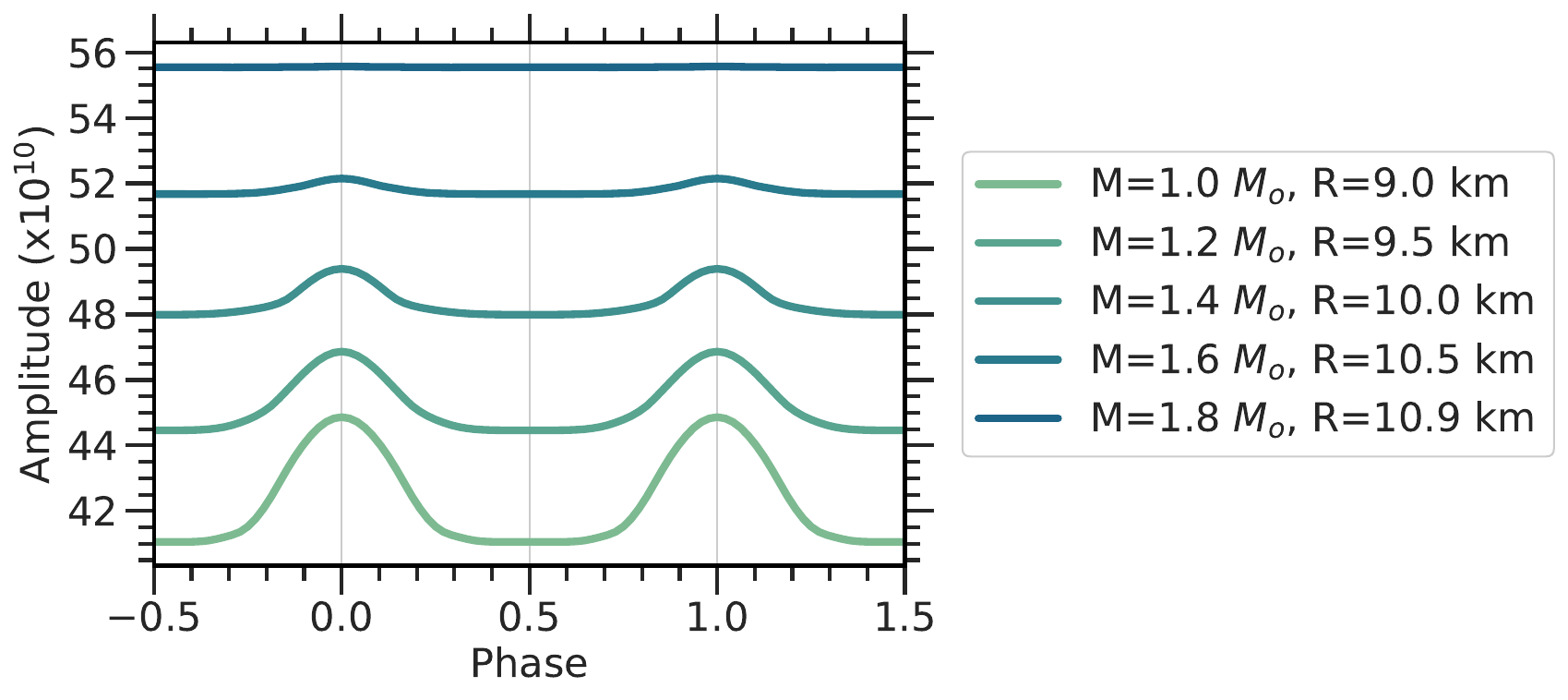}
    \caption{Non-normalized pulse profiles for a uniform isotropic emitter with a limb-darkened circular slab of 1 km$^2$ with composite gravitational light bending for $(i_1=60\textsuperscript{o},~i_2=30\textsuperscript{o})$ and five different values of masses (and radii) of the neutron star. The radii are estimated assuming constant model density $\rho_{\text{10}}$ as calculated for the $M=1.4$~M$_{\odot}$, $R = 10~\text{km}$ case. Fainter curves represent lighter masses (and smaller radii). The spin phase resolution is \texttt{0.001}. A non-pulsed Einstein ring is seen in the most massive $M=1.8$~M$_{\odot}$, $R = 10.9~\text{km}$ limit.}
    \label{fig:vary_M}
\end{figure}

\subsubsection{Non-pulsed limit}

In a limiting case shown in Fig. \ref{fig:vary_M}, this corresponds to $\alpha = \psi = 90\textsuperscript{o}$ i.e. $u$$\sim$$0.5$. Since $u = r_s/R_{\text{NS}}$, this gets seen for $M=1.8$~M$_{\odot}$ in Fig. \ref{fig:vary_M}, producing a seemingly flat \lq pulse' profile despite the rotation of the pulsar. In this limiting case, the pulse profiles seem to tend to the flat profile previously seen in Fig. \ref{fig:isotropic} for two isotropic \textit{point} hotspots \textit{without} gravitational light bending and \textit{without} surface integration. However, such high values of masses and radii are seldom observed and this remains a limiting case\footnote{The masses of Cygnus X-2 and Vela X-1 have been reported near this end i.e. $1.78^{+0.23}_{-0.23}$ M$_{\odot}$ \citep{cygnusx2} and  $1.86^{+0.16}_{-0.16}$ M$_{\odot}$ \citep{velax1, VELAX}. Centrifugal forces and magnetic stresses can help sustain higher neutron star masses against core collapse.}. The constant density approximation, too, can be improved with more pertinent neutron star equations-of-state (See e.g. \cite{2023ApJ...944....7S} for more discussion.). Do note that the overall surface temperature of the neutron star is cooler by a few orders of magnitude and although the emission is negligible compared to the beamed luminosity there are discussions of it producing a non-pulsed background of $70-80\%$.

\subsection{Flux boost and smoothening}

There seems to be an overall increase in the non-pulsed component upon the inclusion of gravitational light bending and integrated emission from a circular surface, in comparison to that of a point hotspot. This is accompanied by a seeming spreading out of emission emanating near the pulse peak because general relativistic bending causes the beamed emission to stretch out across a wider range of angles and a smoothening out of sharp features present in non-integrated profiles.

Surface integration primarily increases the pulsar flux by a few orders of magnitude corresponding to the size of the emission area (in C.G.S. units). In Fig. \ref{fig:surface_integration}, the net flux shows an increase by exactly 10 orders of magnitude from 10$^{28}$ photons/cm$^2$/s/str/keV from the \citeauthor{MeszarosNagel1985b} beaming functions to 10$^{38}$ photons/s/str/keV amplitude units compared to their corresponding point hotspot counterparts, falling in the ballpark of the typical source X-ray pulsar luminosity range of $10^{34} - 10^{38}$ ergs/s.  As expected, in either case, the gravitational bent profiles for a pencil beam are always more luminous at the same spin phases due to the spreading out of the emission beam which results in sampling lower beam angles at the same spin phase and providing additional flux due to bent X-rays from the dark face of the compact star near phase $\phi$ = \texttt{0.25}.

A smoothening out of sharp features otherwise present in non-integrated profiles occurs due to the mixing of X-rays from several infinitesimally small area elements spanning an extended emission region.  This is composite with a further smoothening due to the spreading out of emission near the pulse peak because of composite general relativistic bending which causes the emission beam to stretch out across a wider range of angles. Photons bent toward the observer from the dark face of the neutron star were also computed by composite integrating partially over the second emission area. The composite X-ray photon energies are suitably red-shifted.

\begin{figure}
        \centering
        \includegraphics[trim=5 8 5 5, clip=true, height=0.45\textheight]{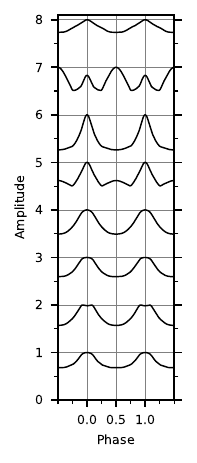}
        \hspace{-.3cm}
        \includegraphics[trim=25 8 5 5, clip=true, height=0.45\textheight]{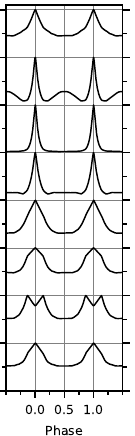}
        \hspace{-.3cm}
        \includegraphics[trim=25 8 5 5, clip=true, height=0.45\textheight]{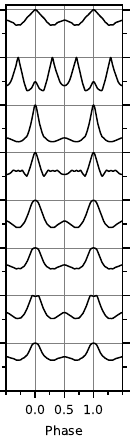}
        \hspace{-.3cm}
        \includegraphics[trim=25 8 5 5, clip=true, height=0.45\textheight]{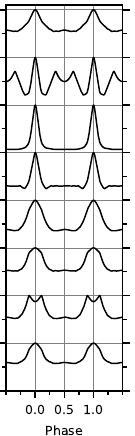}
         \hspace{-.3cm}
        \includegraphics[trim=25 8 5 5, clip=true, height=0.45\textheight]{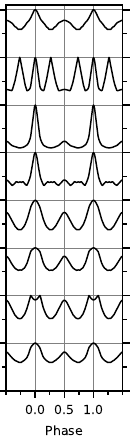}
        \vfill
        \includegraphics[trim=5 8 5 5, clip=true, height=0.45\textheight]{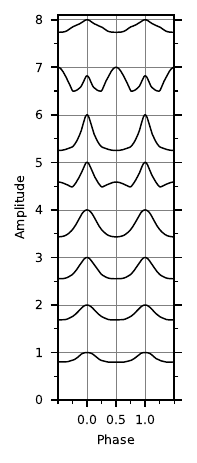}
        \hspace{-.3cm}
        \includegraphics[trim=25 8 5 5, clip=true, height=0.45\textheight]{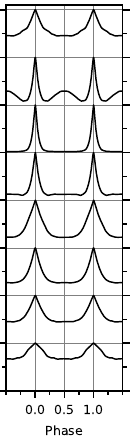}
        \hspace{-.3cm}
        \includegraphics[trim=25 8 5 5, clip=true, height=0.45\textheight]{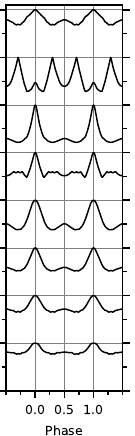}
        \hspace{-.3cm}
        \includegraphics[trim=25 8 5 5, clip=true, height=0.45\textheight]{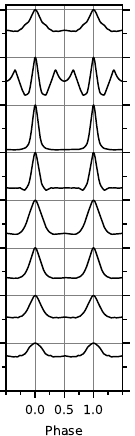}
         \hspace{-.3cm}
        \includegraphics[trim=25 8 5 5, clip=true, height=0.45\textheight]{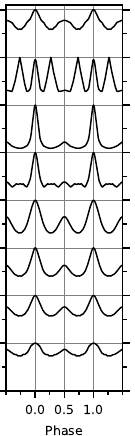}
        \caption{Normalized composite hotspot integrated and gravitationally bent bolometric pulse profiles as in Figs. \ref{fig:pulses_gbl_1} and \ref{fig:pulses_gbl_2} integrating the emission over a circular slab with a radial resolution of 0.01 km and an azimuthal resolution of 10$^{\circ}$ over the hotspot surface (emission area resolution of $<0.001$ km$^2$). The profiles exhibit additional smoothing and reduced dips while remaining consistent with \citeauthor{MeszarosNagel1985b}'s results.}
        \label{fig:pulses_surfint}
\end{figure}

Fig. \ref{fig:compare} shows the non-normalized flux profiles for a shallow slab emission geometry. Surface-integrated profiles shown in red colour are smoother and lie on a scale ten times higher in order of magnitude in the expected range for XRPs. The two profiles in the upper region are broadened due to gravitational bending. These would also correspond to a red-shifted energy. 

\begin{figure}
    \centering
    \includegraphics[width=.4\textwidth, angle=90]{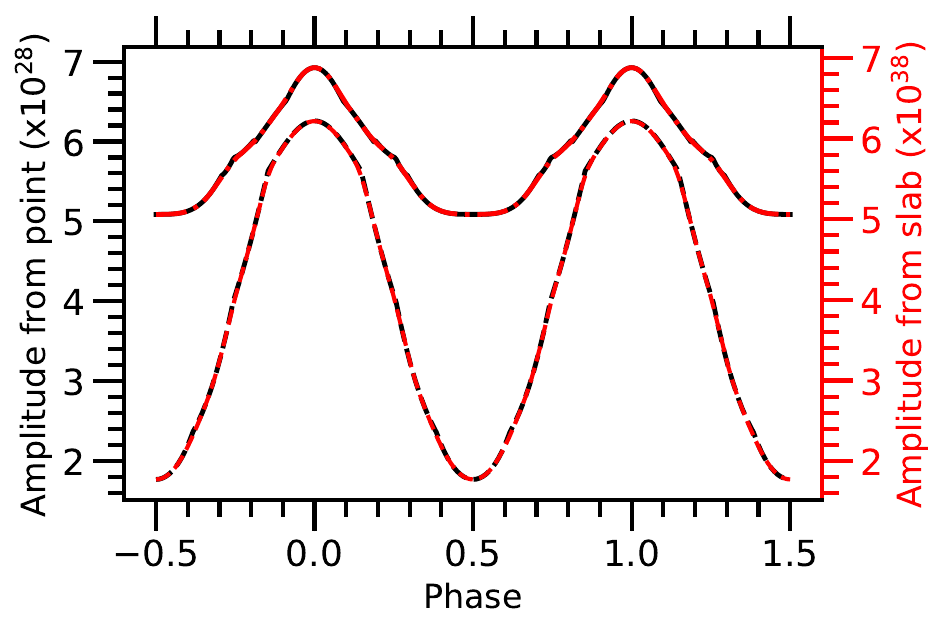}
    \caption{Non-normalized pulse profiles for shallow slab emission geometry with source inclination ($i_1=50\textsuperscript{o}$, $i_2=20\textsuperscript{o}$) at 84.7 keV. The flux amplitude on the left $Y$-axis is in units of photons/cm$^2$/s/str/keV and on the right $Y$-axis is in the units of photons/s/str/keV. The black curves and $Y$-axis denote point hotspots with the dashed black curve representing the model \citeauthor{MeszarosNagel1985b} profile and the solid black curve showing the same with light bending. The red curves and $Y$-axis denote profiles surface-integrated over a model circular slab of 10$^{10}$ cm$^2$ emission area with the dashed red curve representing smoothened profiles without light bending and the dashed-dotted red curve showing smoothened profiles with composite surface integration and light bending. The spin phase resolution is \texttt{0.005}. The curves with GR effects are red-shifted to 70.2 keV.}
    \label{fig:surface_integration}
        \label{fig:compare}
\end{figure}

\subsection{Comparison with a test isotropic beam} \label{app:isotropic}

Pulse profiles for an isotropic hotspot with a uniform beaming function of unity (in arbitrary units) for all X-ray photon energies and beaming angles were obtained in Fig. \ref{fig:isotropic}. The radiation in such a case would be completely unbeamed in the absence of a strong collimating dipolar magnetic field and there would be no preferential direction of photon direction. An example source inclination of ($i_1=60\textsuperscript{o}$, $i_2=30\textsuperscript{o}$) is used for demonstration in this section. 

In the absence of the magnetic field, the pulse fraction would reduce and asymptote to vanishing values for the limiting case of uniform, isotropic emitters considered in previous Sec. \ref{app:limb}, thus presenting flat, non-pulsed X-ray flux profiles to the observer, as though the rotating compact object were a persistent source. The pulse profile is completely flat in panel \texttt{(c)}, appearing the same as the uniform beaming function since the second hotspot comes into view when the first one goes out of sight. Note that there would be a momentary sharp peak up to 2.0 units near the phase $\phi$ = \texttt{0.25} where both hotspots are visible as unit flux would be received at $\alpha=90\textsuperscript{o}$ from both hotspots. However, this instant would be an extremely narrow feature in the pulse profile for point hotspots and hence will remain unresolved for all practical purposes -- yielding completely flat profiles with unit amplitude. Uniform injection assures a minimum flux of 1.0 units from at least one of the two antipodal hotspots being visible at all time instances. 

As seen in panel \texttt{(b)} of Fig. \ref{fig:isotropic}, gravitational light bending results in a peak before spin phase $\phi= \texttt{0.25}$\footnote{As depicted in the diagram, the exact value is $\phi$ = \texttt{0.142} and is determined by the $\psi_{\text{max}}$ parameter which is a function of the compactness parameter $u$. The compactness, in turn, quantifies the strength of the gravitational field of the neutron star, which governs the extent of bending to which photon trajectories are subjected.} due to equal and additional contribution from the other hotspot. Panel \texttt{(a)} shows the composite profile with slab integration. The increase in flux due to an excess contribution from the \lq invisible' hotspot remains at the same phase. However, instead of a sharp increase, the rise is gradual with phase owing to the emission emanating from an extended region rather than a single sharp point. However, as the model emission area of the hotspot is still quite small compared to the total surface area of the neutron star, the slope remains steep. The slope will grow increasingly slanted for greater slab sizes. Since the beaming function is a uniform unity, the maximum flux is slightly below $2 \times 10^{10}$ units but the amplitude has increased by ten orders of magnitude due to a double summation over fluxes from the entire emission area of 1 km$^2$. The minimum value of flux is slightly below $1 \times 10^{10}$ units.

\begin{figure}
  \centering
    \includegraphics[clip, trim = 200 0 330 20, width=.5\textwidth]{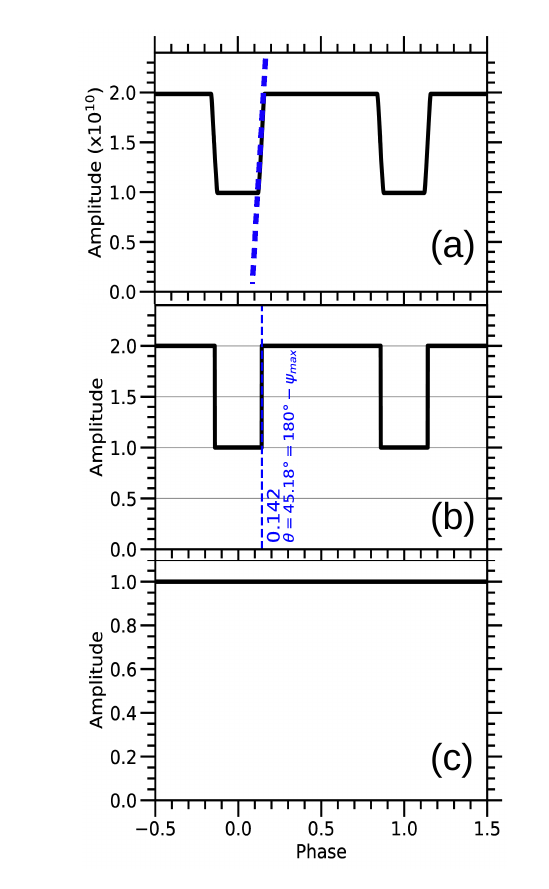}
    \caption{Non-normalized pulse profiles (solid black lines) for uniform injection of an arbitrary X-ray photon energy \texttt{(c)} for a point hotspot, \texttt{(b)} with gravitational light bending and \texttt{(a)} composite with surface integration over a 1 km$^2$ circular slab for $(i_1=60\textsuperscript{o},~i_2=30\textsuperscript{o})$. The spin phase resolution is \texttt{0.001}. Blue-dotted annotations highlight distinct features and provide details.}
    \label{fig:isotropic}
\end{figure}

\section{Conclusions} \label{sec:conclusions}

The paper provides a ground-up description of pulsed observations from accreting strongly magnetized neutron stars, X-ray pulsars using numerical simulations. Existing theoretical models for beamed X-ray emission emanating from such systems are used to develop new, comprehensive numerical tools that simulate model pulse characteristics. This is achieved by incorporating the important physical effects, namely, anisotropic, thermal, extended emission geometric and general relativistic effects, towards modeling observed behaviour, updating any previous studies in this direction with newer theoretical advances reported in the literature. For completeness, several physical possibilities accounting for radiation-matter interactions and geometric configurations are explored.

In Section \ref{subsec:beamed}, model X-ray emission characteristics for a dipolar pulsar magnetic field of strength $\sim$$10^{12}$ G are used to derive the beamed pulse profiles arising from point hotspots by employing direct modeling of spin rotation of a model pulsar. Simulations of phase-resolved pulse profiles that provide a direct prediction of pulse profiles for arbitrary viewing angles for a wide range of emission geometries for a model pulsar and a dipolar magnetic field are obtained. The robust, custom numerical tool developed, the \lq \texttt{AXP4}' can account for several geometries of emission region (and thereby, accretion rates and luminosities) for 8 discrete values of X-ray photon energies spanning the $1-100$ keV range, for all possible inclinations of observer's line-of-sight, [0\textsuperscript{o},~180\textsuperscript{o}) and magnetic axes, [0\textsuperscript{o},~180\textsuperscript{o}) with respect to the pulsar rotation axis. These profiles are further investigated at a high spin phase resolution of the pulsar and with a full range of X-ray photon emission angles (0\textsuperscript{o} --~90\textsuperscript{o}) from the pulsar beam. 

The tool can handle various input parameters corresponding to model physical conditions for binary accreting X-ray pulsars which are prime targets for X-ray observations by building on previous attempts in literature (See e.g. \cite{1981A&A...102...97W, 1991MNRAS.251..203L} and similar works based on \cite{MeszarosNagel1985a}, e.g. \cite{RiffertMeszaros1988, 1993ApJ...406..185R, 1994A&A...286..497P, 1995MNRAS.277.1177L, 2001ApJ...563..289K, 2004ApJ...613..517L, 2007ESASP.622..403T}). 

Known flat-top pulse profiles are first reproduced at a finer phase resolution using model emission beams available in the literature. These are in excellent agreement with standard profiles, thus verifying the operation of the simulator, which becomes the foundation base for building further advanced modules developed sequentially in each section. Model pointed-peak pulse profiles are derived by allowing the model emission functions to extend over the full range of possible source emission angles. 

The effect of general relativistic light bending due to the strong gravitational field around the neutron star which presents brighter pulse profiles to the observer. The \texttt{GR Light Bender} module in Sec. \ref{gbl} captures this effect using  \citeauthor{Beloborodov2002}'s approximation for the curvature of photon trajectories in a Schwarzschild metric, which holds well for the range of distances under consideration in this work and updates previous efforts in this direction. Strong gravity causes the straight beam to broaden out into a larger cone with curved sides which allows flux from lower emission angles to continue remaining visible to the observer at relatively greater rotation spin phases of the pulsar. A key feature of strong bending is the presence of an additional flux component from photon trajectories rooted in the unseen hotspot on the \lq dark' face of the compact star and gravitationally curved towards the other \lq front' face. For cases with sufficiently high gravity, the excess flux can grow to a limiting case of presenting an Einstein ring to the observer. The rotating compact star would be seen as a persistent, non-pulsed emitter. 

Lastly, the escaping X-ray photons undergo gravitational red-shift as shown in Table \ref{tab:z} before reaching the observer, with the effect being increasingly prominent for increasing the energy of photons. This produces a gravitational reddening of the X-ray beam. (The third and last general relativistic effect of a Shapiro time delay between the inter-pulse arrival times for emission from both the hotspots, as well as possible special relativistic effects are significantly applicable only to fast-rotating millisecond pulsars.)

Realistically, the pulsar emission never emanates from a \lq point' hotspot but from an extended area of emission. 
The hotspot area is modeled as a circular slab for a pencil beam of X-ray emission (there are more geometries possible\footnote{These can be readily derived as secondary structures using the primary code for the slabs and columns as the base.} -- see Sec. \ref{future} for further possibilities.). The emission must be integrated over this entire region of emission. A polar Boyer-Lindquist co-ordinate system was used to model a flat circular slab lying flush on the neutron star surface at the polar caps. Computing the total emission over a standard total area of 1 km$^2$ boosts the overall X-ray flux by ten orders of C.G.S. magnitude to fall in the expected typical flux range for binary X-ray pulsars and further smoothens the profiles. Composite gravitationally bent and surface-integrated pulse profiles are further generated for better modeling of realistic scenarios. 

As the pulsar rotates, the inclination of the area, i.e., its projection along the observer's line of sight (more complex for curved surfaces) must be accounted for using a Jacobian, $\mathcal{J}$ which transfers an infinitesimal area element in the source frame to the observer's frame. The curvature of the emission surface presents slightly different emission angles $\alpha$ for a given line of sight $\psi$. Averaging such differential emission over a finite surface on the neutron star surface produces a smoothening of any sharp features present in the profiles. Moreover, any additional contribution from the second antipodal emission area being an extended region of emission is also accounted for. 

\section{Future Scope} \label{future}

Such a cornerstone code presented in this paper for circular slabs could be extended step-wise to accommodate other sub-geometries like multiple ovals \citep{miller2019psr, 2019ApJ...887L..21R} using suitable limits of integration for an elliptical grid and an additive operation for the net emission, crescents \citep{miller2019psr} using a subtractive operation on individual emission from two sets of antipodal hotspots with an inter-slab offset, rings (full, fractional and partial-filled) using suitable limits of integration for the polar grid on the slab, including multiple-temperature components with a suitable modification in the beam input and for non-antipodal (including both in the same rotational sphere) hotspots with unshared parameters \citep{miller2019psr} not excluding the contribution from the second hotspot. The exercise with gravitational light bending and red-shift can be extended to accretion columns as per the prescription by \cite{Falco2016}, accounting for the gravitational gradient along the height of the column. The computed results can be compared with observations obtained from various global X-ray satellites -- both existing and upcoming. The accompanying paper, \cite{2024arXiv240807504S} compares these simulations with Cen X-3 observations analyzed in \cite{2021JApA...42...58S}.

The work carried out in this thesis and the subsequent results presented are polarisation-summed. Given the onset of global X-ray polarimetric space missions opening a new window of high energy polarisation and ushering in an era of X-ray polarisation studies, these results can be resolved into the $X$ and $O$ polarisation modes maintaining a separate treatment for each mode. To this end, a description of how the prescription used by \cite{Meszaros1988} can be customized for the work is outlined in detail in \cite{phdthesisParisee}. Elements from polarised radiative transfer \citep{2018MNRAS.475...43M} and parallel transport and bending of polarised light \citep{2020A&A...641A.126B} would have to be additionally incorporated. Since polarimetric studies aim to resolve the pencil and fan beam degeneracy, alternate theoretical approaches for addressing this issue through the compactness parameter, $u$ are being considered.

The \texttt{AXP4} simulator has been beta-tested and is intended to be made publicly available (open access) after adding some user-friendly layers and along with the requisite documentation\footnote{Subject to academic tenure}. As seen in Fig. \ref{fig:doppler}, the radius of the neutron star has been further parametrized in the same.

\begin{acknowledgments}
The author would like to thank Prof. D. Bhattacharya and Prof. D. Mukherjee for useful discussions, suggestions and ideas. 
Dr. V. Upendran assisted in the initial stages of up-scaling the code in Python \texttt{3}. Prof. G. C. Dewangan, Prof. R. Srianand and Prof. R. Misra provided chronological facilitation for formally reporting the work as a paper, supported by Prof. A N. Ramaprakash. 
\end{acknowledgments}

\facilities{Dell OptiPlex 5060 Desktop\footnote{\url{https://www.dell.com/learn/in/en/inbsd1/shared-content~data-sheets~en/documents~optiplex_5060_spec_sheet.pdf}} with an Intel\textsuperscript{\textregistered} Core\textsuperscript{TM} i7-8700T processor, Ubuntu\textsuperscript{\textregistered} 18.04.3 LTS (64-bit) Operating System, 8 GB DDR4 RAM \& 2 TB HDD.}

\software{NumPy\footnote{\url{https://numpy.org}} \texttt{Ver 1.16.4} \citep{harris}, SciPy\footnote{\url{https://scipy.org}} \texttt{Ver 1.3.0} \citep{virtanen}, Matplotlib\footnote{\url{https://matplotlib.org}} \texttt{Ver 3.1.0} \citep{hunter} packages in Jupyter environment \citep{jupyter} for Python \texttt{3} \citep{rossum},
the SAO/NASA Astrophysics Data System\footnote{\url{https://ui.adsabs.harvard.edu/}}, 
e-Print arXiv\footnote{\url{https://arxiv.org/}}.}

\appendix
\section{High-resolution profiles}
Fig. \ref{fig:resolution} shows the effect of a finer spin phase sampling on resolving sharper features/kinks that might be present in the resultant pulsar profile.
\begin{figure}
    \centering
    \includegraphics[width=.45\linewidth]{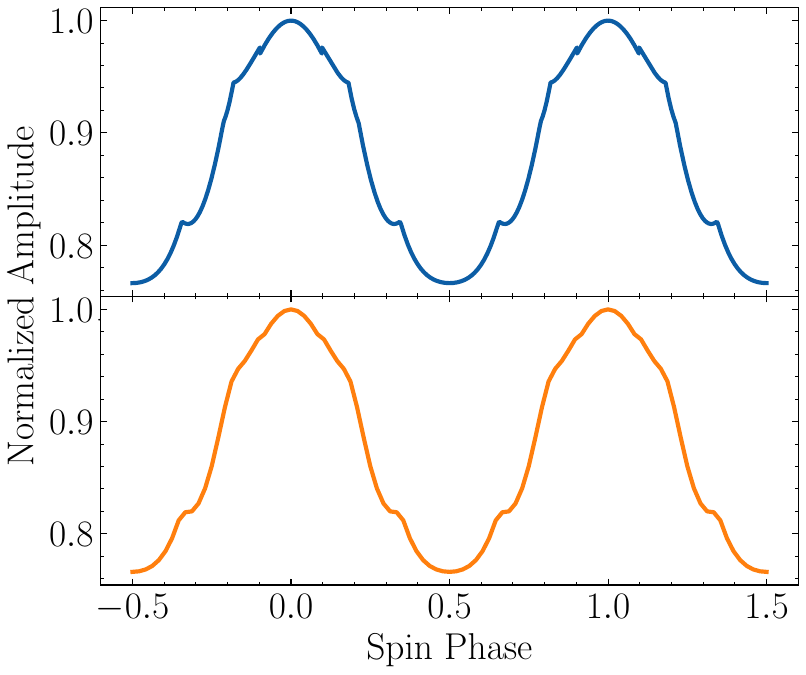}
    \includegraphics[width=.45\linewidth]{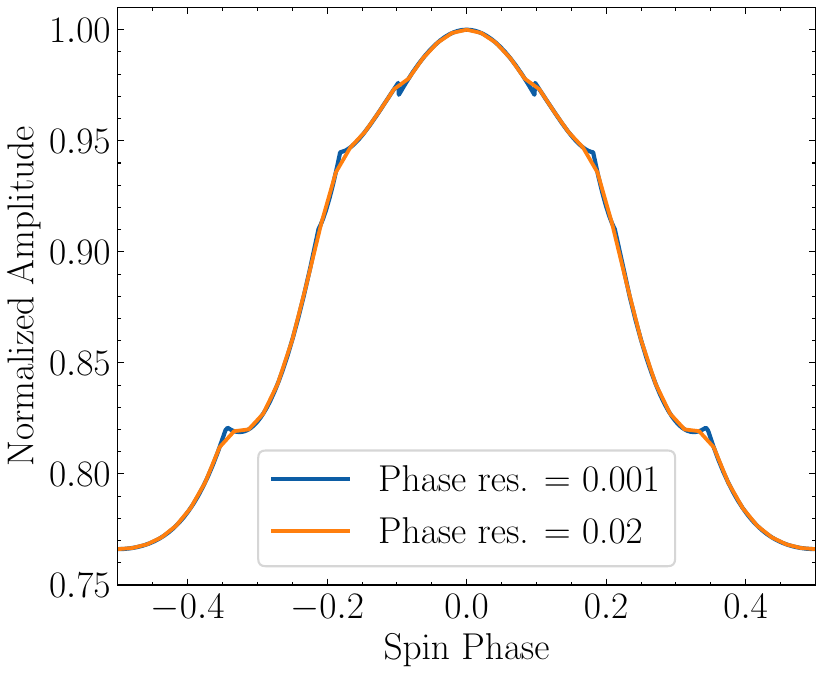}
    \caption{\textit{(left)} Effect of change in spin phase resolution from \texttt{0.02} shown in orange colour to \texttt{0.001} shown in blue colour on the extrapolated gravitationally bent pulse profile including the effect of dark hotspot at gravitationally red-shifted energy of 70.2 keV for a deep slab at ($i_1$ = 60\textsuperscript{o},~$i_2$ = 20\textsuperscript{o}) and \textit{(right)} the same profiles overlaid on top of each other for direct comparison. High-resolution profiles exhibit kinks that get washed out with low-resolution simulations. Surface integration is kept \texttt{OFF} as it smoothens out the kinks.}
    \label{fig:resolution}
\end{figure}

\section{Exact light bending} \label{exact}
The exact formula for the gravitational bending of light has been provided by \cite{Misner:1973prb} which relates $\alpha$ and $\psi$ in a Schwarzschild geometry as,
\begin{equation}
    \psi = \int\limits_{R}^{\infty} \frac{dr}{r^2} \Bigg[ \frac{1}{b^2} - \frac{1}{r^2} \Bigg( 1- \frac{r_{\text{Sch}}}{r} \Bigg) \Bigg]^{-1/2},
\end{equation}
where $b$ is the impact parameter,
\begin{equation}
    b = \frac{R}{\sqrt{1-u}} \sin{\alpha}.
\end{equation}
Fig. \ref{fig:doppler}a shows a comparison between the implementation of \citeauthor{Beloborodov2002}'s approximate gravitational light bending with the exact formula (See Eqs. (A1) and (A2) of \cite{PoutanenBeloborodov2006}) for an \texttt{AXP4} simulated pulse profile. As expected from the justification provided in Sec. \ref{Metho:2}, the two results are coincident.

\section{Special relativistic correction}

\begin{figure}
    \centering
    \includegraphics[width=.45\linewidth]{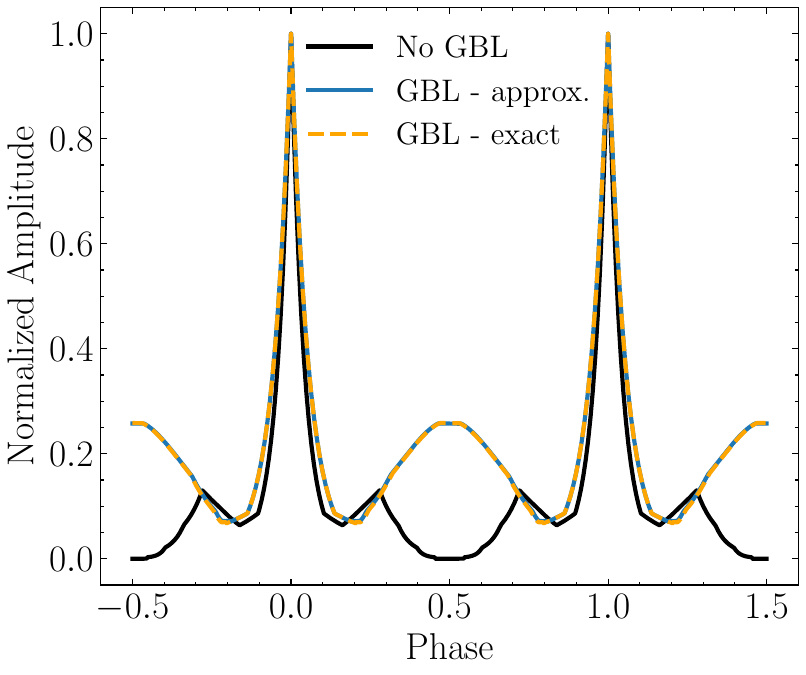} \hfill
    \includegraphics[width=.45\linewidth]{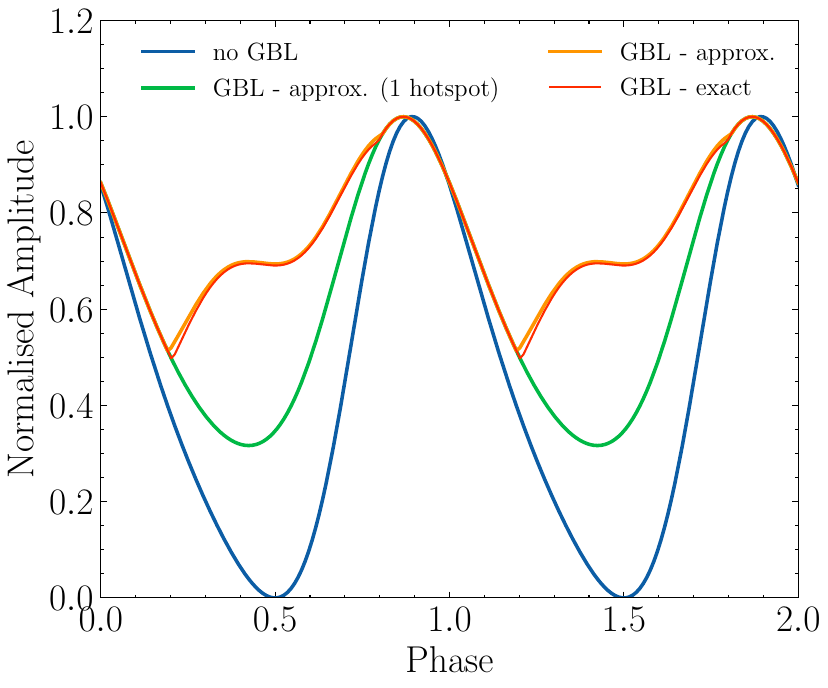}
    \caption{(\textit{left}) Normalized pulse profiles for a model neutron star of 1.4 M$_{\odot}$ mass and 10 km radius inclined at ($i_1$ = 45\textsuperscript{o},~$i_2$ = 45\textsuperscript{o}) with a deep slab observed at the X-ray photon energy of 51.7 keV without gravitational redshift. The graph compares the pulse profiles simulated without the inclusion of gravitational bending of light (GBL) with the corresponding bent profiles simulated using \citeauthor{Beloborodov2002}'s approximation and the exact formula \citep{Misner:1973prb}. There is no visible difference between the latter two. (\textit{right}) Skewed Doppler boosted profiles for two isotropic slab emitters of a fast rotating pulsar with a $600$ Hz spin frequency, 1.4 M$_{\odot}$ mass, $R=2rs=10.336$ km radius and ($i_1$ = 45\textsuperscript{o},~$i_2$ = 45\textsuperscript{o}) inclination consistent with \cite{PoutanenBeloborodov2006}. At this inclination, the single and double hotspot profiles are degenerate without bending (blue curve). The phase axis has been matched to their Fig. 2. When the composite effect of the gravitational bending of light (GBL) is included, the approximate \citep{Beloborodov2002} and the exact formula \citep{Misner:1973prb} remain coincident for a single hotspot (green curve) but become distinguishable in case of two hotspots (yellow and red curves) when the Doppler effect is active.}
    \label{fig:doppler}
\end{figure}

Fig. \ref{fig:doppler}b displays the skewed Doppler boosted profiles for two antipodal isotropic slab emitters of a fast rotating pulsar. The bolometric flux in Eq. (A16) of \cite{PoutanenBeloborodov2006} is reduced to $dF \propto \delta^5 \cos{\alpha}$ in the absence of gravitational bending and surface integration (a typical spectral power law index of $\Gamma=2$ is chosen for accreting millisecond pulsars). This figure can be compared with their Fig. 2 which provides results for only a single hotspot. 

\newpage
\bibliography{sample631.bib}
\bibliographystyle{aasjournal}

\end{document}